\documentclass[runningheads]{llncs}

\makeatletter
\RequirePackage[bookmarks,unicode,colorlinks=true,pdfusetitle]{hyperref}%
   \def\@citecolor{blue}%
   \def\@urlcolor{blue}%
   \def\@linkcolor{blue}%

\def\orcidID#1{\smash{\href{http://orcid.org/#1}{\protect\raisebox{-1.25pt}{\protect\includegraphics{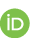}}}}}
\makeatother

\usepackage{graphicx}
\hypersetup{
	pdfauthor={Satoshi Kura}
}

\usepackage{cite}
\usepackage{bussproofs}
\usepackage{amsmath,amssymb}
\usepackage{stmaryrd}
\usepackage{wrapfig}
\usepackage{tikz}
\usetikzlibrary{cd}
\usepackage{mathtools}
\mathtoolsset{showonlyrefs=true}
\usepackage{marvosym}

\usepackage{ifthen}
\newboolean{longversion}
\ifthenelse{\equal{\detokenize{long}}{\jobname}}{
	\setboolean{longversion}{true}
}{
	\setboolean{longversion}{false}
}
\newcommand{\referappendix}[2]{\ifthenelse{\boolean{longversion}}{\S\ref{#1}}{\cite[Appendix {#2}]{arxiv}}}

\newcommand{\Set}{\mathbf{Set}}
\newcommand{\monoid}[1]{\mathbf{#1}}
\newcommand{\theory}[1]{\mathcal{#1}}
\newcommand{\nat}{\mathbb{N}}
\newcommand{\mult}{\mathrel{\otimes}}
\newcommand{\identity}{I}
\newcommand{\comp}{\mathrel{\circ}}
\newcommand{\modelcat}[1]{\mathrm{Mod}_{#1}}
\newcommand{\act}{\mathrel{\circledast}}
\newcommand{\freefunctor}[1]{F^{#1}}
\newcommand{\category}[1]{\mathbf{#1}}

\newcommand{\obj}{\mathrm{ob}}
\newcommand{\rclosed}[2]{\left[{#1}, {#2}\right]}
\newcommand{\lclosed}[2]{\left\llbracket {#1}, {#2} \right\rrbracket}
\newcommand{\tensor}{\mathrel{\otimes}}
\newcommand{\cotensor}{\pitchfork}
\newcommand{\op}{\mathrm{op}}
\newcommand{\twist}[1]{{#1}^{\mathrm{t}}}

\newcommand{\Day}[1]{\check{#1}}
\newcommand{\Daytensor}{\mathrel{\Day{\mult}}}
\newcommand{\equivcat}{\simeq}
\newcommand{\GradedMonad}{\mathbf{GMnd}}
\newcommand{\GradedLawvere}{\mathbf{GLaw}}
\newcommand{\GradedSpec}{\mathbf{GS}}
\newcommand{\initialGLawop}[1]{\mathbf{N}_{#1}}
\newcommand{\initialGLaw}[1]{\initialGLawop{#1}^{\op}}
\newcommand{\theoryfunctor}{\mathbf{Th}}
\newcommand{\MSet}[1]{[\category{#1}, \Set]}
\newcommand{\ordMSet}[1]{\MSet{#1}_0}

\newcommand{\Terms}[3]{T^{#1}_{#2}(#3)}
\newcommand{\barcomp}{\mathrel{\overline{\comp}}}
\newcommand{\KleisliLike}[2]{\widetilde{{#1}_{#2}}}

\spnewtheorem{mytheorem}{Theorem}{\bfseries}{\itshape}
\spnewtheorem{mylemma}[mytheorem]{Lemma}{\bfseries}{\itshape}
\spnewtheorem{myproposition}[mytheorem]{Proposition}{\bfseries}{\itshape}
\spnewtheorem{mysublemma}[mytheorem]{Sublemma}{\bfseries}{\itshape}
\spnewtheorem{mycorollary}[mytheorem]{Corollary}{\bfseries}{\itshape}
\spnewtheorem{myfact}[mytheorem]{Fact}{\bfseries}{\itshape}
\spnewtheorem{mynotation}[mytheorem]{Notation}{\bfseries}{\rmfamily}
\spnewtheorem{myremark}[mytheorem]{Remark}{\bfseries}{\rmfamily}
\spnewtheorem{myexample}[mytheorem]{Example}{\bfseries}{\rmfamily}
\spnewtheorem{myassumption}[mytheorem]{Assumption}{\bfseries}{\rmfamily}
\spnewtheorem{mydefinition}[mytheorem]{Definition}{\bfseries}{\rmfamily}
\spnewtheorem{myrequirements}[mytheorem]{Requirements}{\bfseries}{\rmfamily}
\spnewtheorem{myproblem}[mytheorem]{Problem}{\bfseries}{\rmfamily}
\spnewtheorem{myconjecture}[mytheorem]{Conjecture}{\bfseries}{\rmfamily}

\title{Graded Algebraic Theories}
\author{Satoshi Kura\inst{1,2}\orcidID{0000-0002-3954-8255}\Letter}
\institute{National Institute of Informatics, Tokyo, Japan
\and The Graduate University for Advanced Studies (SOKENDAI), Kanagawa, Japan \\
\email{kura@nii.ac.jp}}

\titlerunning{Graded Algebraic Theories}
\authorrunning{S. Kura}

\begin{document}
\maketitle

\begin{abstract}
	We provide graded extensions of algebraic theories and Lawvere theories that correspond to graded monads.
	We prove that graded algebraic theories, graded Lawvere theories, and finitary graded monads are equivalent via equivalence of categories, which extends the equivalence for monads.
	We also give sums and tensor products of graded algebraic theories to combine computational effects as an example of importing techniques based on algebraic theories to graded monads.
\end{abstract}

\section{Introduction}\label{sec:introduction}

In the field of denotational semantics of programming languages, monads have been used to express computational effects since Moggi's seminal work~\cite{DBLP:conf/lics/Moggi89}. They have many applications from both theoretical and practical points of view.

Monads correspond to \emph{algebraic theories}~\cite{Linton66}.
This correspondence gives natural presentations of many kinds of computational effects by operations and equations~\cite{DBLP:conf/fossacs/PlotkinP02}, which is the basis of algebraic effect~\cite{DBLP:conf/fossacs/PlotkinP01}.
The algebraic perspective of monads also provides ways of combining~\cite{DBLP:journals/tcs/HylandPP06}, reasoning about~\cite{DBLP:conf/lics/PlotkinP08}, and handling computational effects~\cite{DBLP:journals/corr/PlotkinP13}.

\emph{Graded monads}~\cite{smirnov2008graded} are a refinement of monads and defined as a monad-like structure indexed by a monoidal category (or a preordered monoid). The unit and multiplication of graded monads are required to respect the monoidal structure.
This structure enables graded monads to express some kind of ``abstraction'' of effectful computations.
For example, graded monads are used to give denotational semantics of effect systems~\cite{DBLP:conf/popl/Katsumata14}, which are type systems designed to estimate scopes of computational effects caused by programs.

\begin{wrapfigure}{r}{19.6em}
	\centering
	\vspace{-1.5em}
	\hspace*{-4mm}
	\def\defaultHypSeparation{\hskip .1in}
	\AxiomC{$f \in \Sigma_{n, m}$}
	\AxiomC{$t_i \in \Terms{\Sigma}{m'}{X}$ for each $i \in \{1, \dots, n\}$}
	\BinaryInfC{$f(t_1, \dots, t_n) \in \Terms{\Sigma}{m \mult m'}{X}$}
	\scalebox{0.9}{\DisplayProof}
	\caption{A rule of term formation.}\label{fig:term_rule}
	\vspace{-2em}
\end{wrapfigure}
This paper provides a \emph{graded extension of algebraic theories} that corresponds to monads graded by small strict monoidal categories. This generalizes $\nat$-graded theories in~\cite{DBLP:conf/calco/MiliusPS15}.
The main ideas of this extension are the following.
First, we assign to each operation a \emph{grade}, i.e., an object in a monoidal category that represents effects.
Second, our extension provides a mechanism (Fig~\ref{fig:term_rule}) to keep track of effects in the same way as graded monads.
That is, if an operation $f$ with grade $m$ is applied to terms with grade $m'$, then the grade of the whole term is the product $m \mult m'$.

For example, graded algebraic theories enable us to estimate (an overapproximation of) the set of memory locations computations may access.
The side-effects theory~\cite{DBLP:conf/fossacs/PlotkinP02} is given by operations $\mathsf{lookup}_l$ and $\mathsf{update}_{l, v}$ for each location $l \in L$ and value $v \in V$ together with several equations, and each term represents a computation with side-effects.
Since $\mathsf{lookup}_l$ and $\mathsf{update}_{l, v}$ only read from or write to the location $l$, we assign $\{l\} \in \mathbf{2}^L$ as the grade of the operations in the graded version of the side-effects theory where $\mathbf{2}^L$ is the join-semilattice of subsets of locations $L$.
The grade of a term is (an overapproximation of) the set of memory locations the computations may access thanks to the rule in Fig~\ref{fig:term_rule}.

We also provide \emph{graded Lawvere theories} that correspond to graded algebraic theories.
The intuition of a Lawvere theory is a category whose arrows are terms of an algebraic theory.
We use this intuition to define graded Lawvere theories.
In graded algebraic theories, each term has a grade, and substitution of terms must respect the monoidal structure of grades.
To characterize this structure of ``graded'' terms, we consider Lawvere theories enriched in a presheaf category.

Like algebraic theories brought many concepts and techniques to the semantics of computational effects, we expect that the proposed graded algebraic theories will do the same for effect systems.
We look into one example out of such possibilities: combining graded algebraic theories.

The main contributions of this paper are summarized as follows.
\begin{itemize}
	\item We generalize ($\nat$-)graded algebraic theories of~\cite{DBLP:conf/calco/MiliusPS15} to $\category{M}$-graded algebraic theories and also provide $\category{M}$-graded Lawvere theories where $\category{M}$ is a small strict monoidal category.
		We show that there exist translations between these notions and finitary graded monads, which yield equivalences of categories.
	\item We extend sums and tensor products of algebraic theories~\cite{DBLP:journals/tcs/HylandPP06} to graded algebraic theories.
		We define sums in the category of $\category{M}$-graded algebraic theories, and tensor products as an $\category{M} \times \category{M}'$-graded algebraic theory made from an $\category{M}$-graded and an $\category{M}'$-graded algebraic theory.
		We also show a few properties and examples of these constructions.
\end{itemize}

\section{Preliminaries}\label{sec:preliminaries}

\subsection{Enriched Category Theory}
{
\newcommand{\twistmult}{\mathrel{\twist{\mult}}}
We review enriched category theory and introduce notations.
See~\cite{kelly1982basic} for details.

Let $\category{V}_0 = (\category{V}_0, \mult, I)$ be a (not necessarily symmetric) monoidal category.
$\category{V}_0$ is \emph{right closed} if $({-}) \mult X : \category{V}_0 \to \category{V}_0$ has a right adjoint $\rclosed{X}{-}$ for each $X \in \obj \category{V}_0$.
Similarly, $\category{V}_0$ is \emph{left closed} if $X \mult ({-})$ has a right adjoint $\lclosed{X}{-}$ for each $X \in \obj \category{V}_0$.
$\category{V}_0$ is \emph{biclosed} if $\category{V}_0$ is left and right closed.

Let $\twist{\category{V}_0}$ denote the monoidal category $(\category{V}_0, \twistmult, I)$ where $\twistmult$ is defined by $X \twistmult Y \coloneqq Y \mult X$.
Note that $\twist{\category{V}_0}$ is right closed if and only if $\category{V}_0$ is left closed.

We define \emph{$\category{V}_0$-category}, \emph{$\category{V}_0$-functor} and \emph{$\category{V}_0$-natural transformation} as in~\cite{kelly1982basic}.

If $\category{V}_0$ is right closed, then $\category{V}_0$ itself enriches to a $\category{V}_0$-category $\category{V}$ with hom-object given by $\category{V}(X, Y) \coloneqq \rclosed{X}{Y}$.
We use the subscript $({-})_0$ to distinguish the enriched category $\category{V}$ from its underlying category $\category{V}_0$.

Assume that $\category{V}_0$ is biclosed and let $\category{A}$ be a $\category{V}_0$-category.
The \emph{opposite category} $\category{A}^{\op}$ is the $\twist{\category{V}_0}$-category defined by $\category{A}^{\op}(X, Y) = \category{A}(Y, X)$.
For any $X \in \obj \category{A}$, $\category{A}(X, {-}) : \category{A} \to \category{V}_0$ is a $\category{V}_0$-functor where $\category{A}(X, {-})_{Y, Z} : \category{A}(Y, Z) \to [\category{A}(X, Y), \category{A}(X, Z)]$ is defined by transposing the composition law $\overline{\comp}$ of $\category{A}$.
A $\twist{\category{V}_0}$-functor $\category{A}({-}, X)$ is defined by $\category{A}^{\op}(X, {-}) : \category{A}^{\op} \to \twist{\category{V}_0}$.

Let $\category{A}$ be a $\category{V}_0$-category.
For each $X \in \category{V}_0$ and $C \in \category{A}$, a \emph{tensor} $X \tensor C$ is an object in $\category{A}$ together with a counit morphism $\nu : X \to \category{A}(C, X \tensor C)$ such that a $\category{V}_0$-natural transformation
$\category{A}(X \tensor C, {-}) \to \lclosed{X}{\category{A}(C, {-})}$
obtained by transposing $(\overline{\comp}) \comp (\category{A}(X \tensor C, B) \mult \nu)$ is isomorphic where $\overline{\comp}$ is the composition in the $\category{V}_0$-category $\category{A}$.
A \emph{cotensor} $X \cotensor C$ is a tensor in $\category{A}^{\op}$.
For example, if $\category{V}_0 = \Set$, then tensors $X \tensor C$ are copowers $X \cdot C$, and cotensors $X \cotensor C$ are powers $C^X$.

A $\category{V}_0$-functor $F : \category{A} \to \category{B}$ is said to preserve a tensor $X \tensor C$ if $F_{C, X \tensor C} \comp \nu : X \to \category{B}(F C, F(X \tensor C))$
is again a counit morphism.
$F$ preserves cotensors if $F^{\op}$ preserves tensors.

Let $\Phi$ be a collection of objects in $\category{V}_0$.
A $\category{V}_0$-functor $F : \category{A} \to \category{B}$ is said to preserve $\Phi$-(co)tensors if $F$ preserves (co)tensors of the form $X \tensor C$ ($X \cotensor C$) for each $X \in \Phi$ and $C \in \obj \category{A}$.
}

\subsection{Graded Monads}
We review the notion of graded monad in~\cite{DBLP:conf/popl/Katsumata14,DBLP:conf/fossacs/FujiiKM16}, and then define the category $\GradedMonad_{\category{M}}$ of finitary $\category{M}$-graded monads.
Throughout this section, we fix a small strict monoidal category $\category{M} = (\category{M}, {\mult}, \identity)$.

\begin{mydefinition}[graded monads]
	An \emph{$\category{M}$-graded monad} on $\category{C}$ is a lax monoidal functor $\category{M} \to [\category{C}, \category{C}]$ where $[\category{C}, \category{C}]$ is a monoidal category with composition as multiplication.
	That is, an $\category{M}$-graded monad is a tuple $(\ast, \eta, \mu)$ of a functor $\ast : \category{M} \times \category{C} \to \category{C}$ and natural transformations $\eta_X : X \to \identity \ast X$ and $\mu_{m_1, m_2, X} : m_1 \ast (m_2 \ast X) \to (m_1 \mult m_2) \ast X$ such that the following diagrams commute.
	\begin{center}
		\begin{tikzcd}[column sep=small, row sep=small]
			m \ast X \ar[r, "\eta"] \ar[d, swap, "m \ast \eta"] \ar[rd, equal]& \identity \ast (m \ast X) \ar[d, "\mu"] \\
			m \ast (\identity \ast X) \ar[r, swap, "\mu"] & m \ast X
		\end{tikzcd}
		\begin{tikzcd}[row sep=small]
			m_1 {\ast} (m_2 {\ast} (m_3 {\ast} X)) \ar[r, "m_1 \ast \mu"] \ar[d, swap, "\mu"] & m_1 {\ast} ((m_2 {\mult} m_3) {\ast} X) \ar[d, "\mu"] \\
			(m_1 {\mult} m_2) {\ast} (m_3 {\ast} X) \ar[r, swap, "\mu"] & (m_1 {\mult} m_2 {\mult} m_3) {\ast} X
		\end{tikzcd}
	\end{center}

	A \emph{morphism of $\category{M}$-graded monad} is a monoidal natural transformation $\alpha : (\ast, \eta, \mu) \to (\ast', \eta', \mu')$, i.e.\ a natural transformation $\alpha : \ast \to \ast'$ that is compatible with $\eta$ and $\mu$.
\end{mydefinition}

An intuition of graded monads is a refinement of monads: $m \ast X$ is a computation whose scope of effect is indicated by $m$ and whose result is in $X$.
The monoidal category $\category{M}$ defines the granularity of the refinement, and a $\mathbf{1}$-graded monad is just an ordinary monad.
Note that we do not assume that $\category{M}$ is symmetric because some of graded monads in~\cite{DBLP:conf/popl/Katsumata14} require $\category{M}$ to be nonsymmetric.
We also deal with such a nonsymmetric case in Example~\ref{ex:theory_graded_exception_monad}.

A \emph{finitary functor} is a functor that preserves filtered colimits.
In this paper, we focus on finitary graded monads on $\Set$.
\begin{mydefinition}\label{def:finitary_graded_monad}
	A \emph{finitary $\category{M}$-graded monad on $\Set$} is a lax monoidal functor $\category{M} \to [\Set, \Set]_f$ where $[\Set, \Set]_f$ denotes the full subcategory of $[\Set, \Set]$ on finitary functors.
	Let $\GradedMonad_{\category{M}}$ denote the category of finitary $\category{M}$-graded monads and monoidal natural transformations between them.
\end{mydefinition}

A morphism in $\GradedMonad_{\category{M}}$ is determined by the restriction to $\aleph_0 \subseteq \Set$ where $\aleph_0$ is the full subcategory of $\Set$ on natural numbers.
\begin{mylemma}\label{lem:finitary_graded_monad_morphism}
Let $T = (\ast, \eta, \mu)$ and $T' = (\ast', \eta', \mu')$ be finitary $\category{M}$-graded monads.
There exists one-to-one correspondence between the following.
\begin{enumerate}
	\item Morphisms $\alpha : T \to T'$.
	\item Natural transformations $\beta : {\ast} \comp (\category{M} \times i) \to {\ast'} \comp (\category{M} \times i)$ (where $i : \aleph_0 \to \Set$ is the inclusion functor) such that the following diagrams commute for each $n, n' \in \aleph_0$, $m_1, m_2 \in \category{M}$ and $f : n \to m_2 \ast n'$.
	\begin{center}
	\begin{tikzcd}
	n \ar[r, "\eta_n"] \ar[rd, swap, "\eta'_n"] & \identity \ast n \ar[d, "\beta"] \\
	& \identity \ast' n
	\end{tikzcd}
	\begin{tikzpicture}[baseline= (a).base]
	\node[scale=.8] (a) at (0,0){
		\begin{tikzcd}[row sep=small]
		m_1 \ast n \ar[r, "\beta"] \ar[dd, "m_1 \ast f"] & m_1 \ast' n \ar[d, "m_1 \ast' f"] \\
		& m_1 \ast' (m_2 \ast n') \ar[d, "m_1 \ast' \beta"] \\
		m_1 \ast (m_2 \ast n') \ar[d, "\mu"] & m_1 \ast' (m_2 \ast' n') \ar[d, "\mu'"] \\
		(m_1 \mult m_2) \ast n' \ar[r, "\beta"] & (m_1 \mult m_2) \ast' n'
		\end{tikzcd}
	};
	\end{tikzpicture}
	\end{center}
\end{enumerate}
\end{mylemma}
\begin{proof}
	By the equivalence $[\Set, \Set]_f \equivcat [\aleph_0, \Set]$ induced by restriction and the left Kan extension along the inclusion $i : \aleph_0 \to \Set$.
	\qed
\end{proof}

\subsection{Day Convolution}\label{sec:prelim_day_conv}
We describe a monoidal biclosed structure on the (covariant) presheaf category $\ordMSet{M}$ where $\category{M} = (\category{M}, {\mult}, \identity)$ is a small monoidal category~\cite{DayClosedCategories}.
Here, we use the subscript $({-})_0$ to indicate that $\ordMSet{M}$ is an ordinary (not enriched) category since we also use the enriched version $\MSet{M}$ later.

The \emph{external tensor product} $F \boxtimes G : \category{M} \times \category{M} \to \Set$ is defined by $(F \boxtimes G)(m_1, m_2) = F m_1 \times G m_2$ for any $F, G : \category{M} \to \Set$.
\begin{mydefinition}
Let $F, G : \category{M} \to \Set$ be functors.
The \emph{Day tensor product} $F \Daytensor G : \category{M} \to \Set$ is the left Kan extension $\mathrm{Lan}_{\mult}(F \boxtimes G)$ of the external tensor product $F \boxtimes G : \category{M} \times \category{M} \to \Set$ along the tensor product ${\mult} : \category{M} \times \category{M} \to \category{M}$.
\end{mydefinition}
Note that a natural transformation $\overline{\theta} : F \Daytensor G \to H$ is equivalent to a natural transformation $\theta_{m_1, m_2} : F m_1 \times G m_2 \to H (m_1 \mult m_2)$ by the universal property.

The Day convolution induces a monoidal biclosed structure in $\ordMSet{M}$~\cite{DayClosedCategories}.
\begin{myproposition}
	The Day tensor product makes $(\ordMSet{M}, {\Daytensor}, y(\identity))$ a monoidal biclosed category where $y : \category{M}^{\op} \to \ordMSet{M}$ is the Yoneda embedding $y(m) \coloneqq \category{M}(m, {-})$.
\qed
\end{myproposition}

The left and the right closed structure are given by $\lclosed{F}{G} m = \ordMSet{M}(F, \allowbreak G(m {\mult} {-}))$ and $\rclosed{F}{G} m = \ordMSet{M}(F, G({-} {\mult} m))$ for each $m \in \category{M}$, respectively.

Note that since we do not assume $\category{M}$ to be symmetric, neither is $\ordMSet{M}$.
Note also that the twisting and the above construction commute: there is an isomorphism $\twist{\ordMSet{M}} \cong \ordMSet{\twist{M}}$ of monoidal categories.

\subsection{Categories Enriched in a Presheaf Category}
We rephrase the definitions of $\ordMSet{M}$-enriched category, functor and natural transformation in elementary terms.
An $\ordMSet{M}$-category is, so to say, an ``$\category{M}$-graded'' category: each morphism has a grade $m \in \obj \category{M}$ and the grade of the composite of two morphisms with grades $m$ and $m'$ is the product $m \mult m'$ of the grades of each morphism.
Likewise, $\ordMSet{M}$-functors and $\ordMSet{M}$-natural transformations can be also understood as an ``$\category{M}$-graded'' version of ordinary functors and natural transformations.
Specifically, the following lemma holds~\cite{DBLP:conf/popl/CurienFM16}.

\begin{mylemma}\label{lem:M_Set_category}
There is a one-to-one correspondence between (1) an $\ordMSet{M}$-category $\category{C}$ and (2) the following data satisfying the following conditions.
\begin{itemize}
\item A class of objects $\mathrm{ob}\category{C}$.
\item For each $X, Y \in \mathrm{ob}\category{C}$, a hom objects $\category{C}(X, Y) \in \ordMSet{M}$.
\item For each $X \in \mathrm{ob}\category{C}$, an element $1_X \in \category{C}(X, X) \identity$.
\item For each $X, Y, Z \in \mathrm{ob}\category{C}$, a family of morphisms 
$\big({\comp}_{m_1, m_2} : \category{C}(Y, Z) m_1 \times \category{C}(X, Y) m_2 \to \category{C}(X, Z) (m_1 \mult m_2)\big)_{m_1, m_2 \in M}$
which is natural in $m_1$ and $m_2$.
The subscripts $m_1$ and $m_2$ are often omitted.
\end{itemize}
These data must satisfy the identity law $1_Y \comp f = f = f \comp 1_X$ for each $f \in \category{C}(X, Y) m$ and the associativity $(h \comp g) \comp f = h \comp (g \comp f)$ for each $f \in \category{C}(X, Y) m_1$, $g \in \category{C}(Y, Z) m_2$ and $h \in \category{C}(Z, W) m_3$.
\end{mylemma}
\begin{proof}
	The identity $\overline{1_X} : y(\identity) \to \category{C}(X, X)$ in $\category{C}$ corresponds to $1_X \in \category{C}(X, X) \identity$ by the Yoneda lemma, and the composition ${\barcomp} : \category{C}(Y, Z) \Daytensor \category{C}(X, Y) \to \category{C}(X, Z)$ in $\category{C}$ corresponds to the natural transformation ${\comp}_{m_1, m_2} : \category{C}(Y, Z) m_1 \times \category{C}(X, Y) m_2 \allowbreak \to \category{C}(X, Z) (m_1 \mult m_2)$ by the universal property of the Day convolution.
	The rest of the proof is easy.
	\qed
\end{proof}

An $\ordMSet{M}$-functor $F : \category{C} \to \category{D}$ consists of a mapping $X \mapsto FX$ and a natural transformation $F_{X, Y} : \category{C}(X, Y) \to \category{D}(FX, FY)$ (for each $X, Y$) that preserves identities and compositions of morphisms.
An $\ordMSet{M}$-natural transformation $\overline{\alpha} : F \to G$ is a family of elements $\big( \alpha_X \in \category{D}(F X, G X) \identity \big)_{X \in \mathrm{ob}(C)}$ that satisfies $\alpha_Y \comp F f = G f \comp \alpha_X$ for each $f \in \category{C}(X, Y) m$.
Vertical and horizontal compositions of $\ordMSet{M}$-natural transformations are defined as expected.

We introduce a useful construction of $\twist{\ordMSet{M}}$-categories.
Given an $\category{M}$-graded monad (in other words, a lax left $\category{M}$-action) on $\category{C}$, we can define an $\twist{\ordMSet{M}}$-enriched category as follows.
\begin{mydefinition}\label{def:enrichment_by_action}
Let $T = (\ast, \eta, \mu)$ be an $\category{M}$-graded monad on $\category{C}$.
An $\twist{\ordMSet{M}}$-category $\KleisliLike{\category{C}}{T}$ is defined by $\obj{\KleisliLike{\category{C}}{T}} \coloneqq \obj{\category{C}}$ and $\KleisliLike{\category{C}}{T}(X, Y) m \coloneqq \category{C}(X, m \ast Y)$.
The identity morphisms are the unit morphisms $\eta_X \in \KleisliLike{\category{C}}{T}(X, X) I$, and the composite of $f \in \KleisliLike{\category{C}}{T}(Y, Z) m$ and $g \in \KleisliLike{\category{C}}{T}(X, Y) m'$ is $\mu \comp (m \ast g) \comp f$.
\end{mydefinition}

The definition of $\KleisliLike{\category{C}}{T}$ is similar to the definition of the Kleisli categories for ordinary monads.
Actually, $\KleisliLike{\category{C}}{T}$ can be constructed via the Kleisli category $\category{C}_T$ for the graded monad $T$ presented in~\cite{DBLP:conf/fossacs/FujiiKM16} (although $\category{C}_T$ itself is not enriched).
This can be observed by $\category{C}_T((I, X), (m, Y)) \cong \KleisliLike{\category{C}}{T}(X, Y) m$.

\section{Graded Algebraic Theories}\label{sec:graded_algebraic_theory}
We explain a framework of universal algebra for graded monads, which is a natural extension of~\cite{smirnov2008graded,DBLP:conf/calco/MiliusPS15}.
The key idea of this framework is that each term is associated with not only an arity but also a ``grade'', which is represented by an object in a monoidal category $\category{M}$.
We also add coercion construct for terms that changes the grade of terms along a morphism of the monoidal category $\category{M}$.
Then, a mapping that takes $m \in \category{M}$ and a set of variables $X$ and returns the set of terms with grade $m$ (modulo the equational axioms) yields a graded monad.

We fix a small strict monoidal category $\monoid{M} = (\category{M}, {\mult}, \identity)$ throughout this section.
We sometimes identify $n \in \nat$ with $\{ 1, \dots, n \}$, or $\{ x_1, \dots, x_n \}$ if it is used as a set of variables.

\subsection{Equational Logic}
A \emph{signature} is a family of sets of symbols $\Sigma = (\Sigma_{n, m})_{n \in \nat, m \in \category{M}}$.
An element $f \in \Sigma_{n, m}$ is called an operation with arity $n$ and grade $m$.
%
We define a sufficient structure to interpret operations in a category $\category{C}$ as follows.
\begin{mydefinition}
	\emph{$\category{M}$-model condition} is defined by the following conditions on a tuple $(\category{C}, ({\act}, \eta^{\act}, \mu^{\act}))$.
	\begin{itemize}
		\item $\category{C}$ is a category with finite power.
		\item $(\act, \eta^{\act}, \mu^{\act})$ is a strong $\twist{\category{M}}$-action (i.e.\ an $\twist{\category{M}}$-graded monad whose unit and multiplication are invertible).
		\item For each $m \in \category{M}$, $m \act ({-})$ preserves finite powers: $m \act c^n \cong (m \act c)^n$.
	\end{itemize}
\end{mydefinition}
\begin{myexample}\label{example:functor_cat_right_action}
	If $\category{A}$ is a category with finite powers, then the functor category $[\category{M}, \category{A}]$ has strong $\twist{\category{M}}$-action defined by $m \act F \coloneqq F(m \mult ({-}))$ and satisfies $\category{M}$-model condition.
	Especially, $\ordMSet{M}$ satisfies $\category{M}$-model condition.
\end{myexample}

A \emph{model} $A = (A, |\cdot|^A)$ of $\Sigma$ in a category $\category{C}$ satisfying $\category{M}$-model condition consists of an object $A \in \category{C}$ and an interpretation $|f|^A : A^n \to m \act A$ for each $f \in \Sigma_{n, m}$.
A \emph{homomorphism} $\alpha : A \to B$ between two models $A, B$ is a morphism $\alpha : A \to B$ in $\category{C}$ such that $(m \act \alpha) \comp |f|^A = |f|^B \comp \alpha^n$ for each $f \in \Sigma_{n, m}$.

\begin{mydefinition}
	Let $X$ be a set of variables.
	The set of ($\category{M}$-graded) $\Sigma$-terms $\Terms{\Sigma}{m}{X}$ for each $m \in \category{M}$ is defined inductively as follows.
\begin{center}
	\small
	\AxiomC{$x \in X$}
	\UnaryInfC{$x \in \Terms{\Sigma}{\identity}{X}$}
	\scalebox{0.9}{\DisplayProof}
	\AxiomC{$t \in \Terms{\Sigma}{m}{X}$}
	\AxiomC{$w : m \to m'$}
	\BinaryInfC{$c_{w}(t) \in \Terms{\Sigma}{m'}{X}$}
	\scalebox{0.9}{\DisplayProof}
	\AxiomC{$f \in \Sigma_{n, m}$}
	\AxiomC{$\forall i \in \{1, \dots, n\},\ t_i \in \Terms{\Sigma}{m'}{X}$}
	\BinaryInfC{$f(t_1, \dots, t_n) \in \Terms{\Sigma}{m \mult m'}{X}$}
	\scalebox{0.9}{\DisplayProof}
\end{center}
\end{mydefinition}
That is, we build $\Sigma$-terms from variables by applying operations in $\Sigma$ and coercions $c_w$ while keeping track of the grade of terms.
When applying operations, we sometimes write $f(\lambda i \in n. t_i)$ or $f(\lambda i. t_i)$ instead of $f(t_1, \dots, t_n)$.

\begin{mydefinition}\label{def:term_interpretation}
	Let $A$ be a model of a signature $\Sigma$.
	For each $m \in \category{M}$ and $s \in \Terms{\Sigma}{m}{n}$, the \emph{interpretation} $|s|^A : A^n \to m \act A$ is defined as follows.
\begin{itemize}
	\item For any variable $x_i$, $|x_i|^A = \eta^{\act} \comp \pi_i$ where $\pi_i : A^n \to A$ is the $i$-th projection.
	\item For each $w : m' \to m$ and $s \in \Terms{\Sigma}{m'}{\{x_1, \dots, x_n\}}$, $|c_{w}(s)|^A = (w \act A) \comp |s|^A$.
	\item If $f \in \Sigma_{k, m'}$ and $t_i \in \Terms{\Sigma}{m''}{\{x_1, \dots, x_n\}}$ for each $i \in \{ 1, \dots, k \}$, then $|f(t_1, \dots, t_k)|^A$ is defined by the following composite.

	\noindent
	\begin{tikzcd}
		A^n \ar[r, "{\langle |t_1|, \dots, |t_k| \rangle}"] &[20pt] (m'' {\act} A)^k \ar[r, "\cong"] &[-15pt] m'' {\act} A^k \ar[r, "m'' {\act} |f|"] & m'' {\act} (m' {\act} A) \ar[r, "\mu"] &[-15pt] (m' {\mult} m'') {\act} A
	\end{tikzcd}
\end{itemize}
\end{mydefinition}
When we interpret a term $t \in \Terms{\Sigma}{m}{X}$, we need to pick a finite set $n$ such that $\mathrm{fv}(t) \subseteq n \subseteq X$ where $\mathrm{fv}(t)$ is the set of free variables in $t$, but the choice of the finite set does not matter when we consider only equality of interpretations by the following fact.
If $\sigma : n \to n'$ is a renaming of variables and $\overline{\sigma} : \Terms{\Sigma}{m}{n} \to \Terms{\Sigma}{m}{n'}$ is a mapping induced by the renaming $\sigma$, then for each $t \in \Terms{\Sigma}{m}{n}$, $|\overline{\sigma}(t)|^A = |t|^A \comp A^{\sigma}$, which implies that equality of the interpretations of two terms $s, t$ is preserved by renaming: $|s| = |t|$ implies $|\overline{\sigma}(s)| = |\overline{\sigma}(s)|$.

An \emph{equational axiom} is a family of sets $E = (E_m)_{m \in \category{M}}$ where $E_m$ is a set of pairs of terms in $\Terms{\Sigma}{m}{X}$.
We sometimes identify $E$ with its union $\bigcup_{m \in \category{M}} E_m$.
A \emph{presentation of an $\monoid{M}$-graded algebraic theory} (or an \emph{$\monoid{M}$-graded algebraic theory}) is a pair $\theory{T} = (\Sigma, E)$ of a signature and an equational axiom.
A \emph{model} $A$ of $(\Sigma, E)$ is a model of $\Sigma$ that satisfies $|s|^A = |t|^A$ for each $(s = t) \in E$.
Let $\modelcat{\theory{T}}(\category{C})$ denote the category of models of $\theory{T}$ in $\category{C}$ and homomorphisms between them.

To obtain a graded monad on $\Set$ from $\theory{T}$, we need a strict left action of $\category{M}$ on $\modelcat{\theory{T}}(\ordMSet{M})$ and an adjunction between $\modelcat{\theory{T}}(\ordMSet{M})$ and $\Set$.
The former is defined by the following, while the latter is described in~\S\ref{subsec:free_model}.

\begin{mylemma}\label{lem:model_action}
	Let $\category{C}$ be a category satisfying $\category{M}_1 \times \category{M}_2$-model condition.
	If $\theory{T}$ is an $\category{M}_1$-graded algebraic theory, then $\category{C}$ satisfies $\category{M}_1$-model condition and $\modelcat{\theory{T}}(\category{C})$ satisfies $\category{M}_2$-model condition.
\end{mylemma}
\begin{proof}
	An $\twist{\category{M}}_1$-action on $\category{C}$ is obtained by the composition of $\twist{\category{M}}_1 \times \twist{\category{M}}_2$-action and the strong monoidal functor $\twist{\category{M}}_1 \to \twist{\category{M}}_1 \times \twist{\category{M}}_2$ defined by $m \mapsto (m, I)$.
	Finite powers and an $\twist{\category{M}}_2$-action for $\modelcat{\theory{T}}(\category{C})$ are induced by those for $\category{C}$.
	\qed
\end{proof}

\begin{mycorollary}\label{cor:MSet_left_action}
	$\modelcat{\theory{T}}(\ordMSet{M})$ has an $\category{M}$-action, which is given by the precomposition of $m \mult ({-})$ like the $\category{M}$-action of Example~\ref{example:functor_cat_right_action}.
\end{mycorollary}
\begin{proof}
	$\ordMSet{M}$ has $\twist{\category{M}} \times \category{M}$-action defined by $(m_1, m_2) * F = F (m_1 \mult ({-}) \mult m_2)$.
	Thus, $\category{M}$-action for $\modelcat{\theory{T}}(\ordMSet{M})$ is obtained by Lemma~\ref{lem:model_action}.
	\qed
\end{proof}

Substitution $s[t_1/x_1, \dots, t_k/x_k]$ for $\category{M}$-graded $\Sigma$-terms can be defined as usual, but we have to take care of grades: given $s \in \Terms{\Sigma}{m}{k}$ and $t_1, \dots, t_k \in \Terms{\Sigma}{m'}{n}$, the substitution $s[t_1/x_1, \dots, t_k/x_k]$ is defined as a term in $\Terms{\Sigma}{m \mult m'}{n}$.

We obtain an equational logic for graded theories by adding some additional rules to the usual equational logic.
\begin{mydefinition}\label{def:equational_logic}
	The entailment relation $\theory{T} \vdash s = t$ (where $s, t \in T_m(X)$) for an $\category{M}$-graded theory $\theory{T}$ is defined by adding the following rules to the standard rules i.e.\ reflexivity, symmetry, transitivity, congruence, substitution and axiom in $E$ (see e.g.\ \cite{sankappanavar1981course} for the standard rules of equational logic).
\begin{center}
	\small
	\AxiomC{$s, t \in \Terms{\Sigma}{m}{X}$}
	\AxiomC{$\theory{T} \vdash s = t$}
	\AxiomC{$w : m \to m'$}
	\TrinaryInfC{$\theory{T} \vdash c_{w} (s) = c_{w}(t)$}
	\scalebox{0.9}{\DisplayProof}
	\hspace{2em}
	\AxiomC{$t \in \Terms{\Sigma}{m}{X}$}
	\UnaryInfC{$\theory{T} \vdash c_{1_m}(t) = t$}
	\scalebox{0.9}{\DisplayProof}\\[5pt]
	\AxiomC{$t \in \Terms{\Sigma}{m}{X}$}
	\AxiomC{$w : m \to m'$}
	\AxiomC{$w' : m' \to m''$}
	\TrinaryInfC{$\theory{T} \vdash c_{w'}(c_{w}(t)) = c_{w' \comp w}(t)$}
	\scalebox{0.9}{\DisplayProof}\\[5pt]
	\AxiomC{$f \in \Sigma_{n, m}$}
	\AxiomC{$t_i \in \Terms{\Sigma}{m'}{X}$ for each $i \in \{ 1, \dots, n\}$}
	\AxiomC{$w : m' \to m''$}
	\TrinaryInfC{$\theory{T} \vdash f(c_{w}(t_1), \dots, c_{w}(t_n)) = c_{m \mult w}(f(t_1, \dots, t_n))$}
	\scalebox{0.9}{\DisplayProof}
\end{center}
\end{mydefinition}

\begin{mydefinition}
	Given a model $A$ of $\theory{T}$, we denote $A \Vdash s = t$ if $s, t \in \Terms{\Sigma}{m}{n}$ (for some $n$) and $|s|^A = |t|^A$.
	If $\category{C}$ is a category satisfying $\category{M}$-model condition, we denote $\theory{T}, \category{C} \Vdash s = t$ if $A \Vdash s = t$ for any model $A$ of $\theory{T}$ in $\category{C}$.
\end{mydefinition}

It is easy to verify that the equational logic in Definition~\ref{def:equational_logic} is sound.
\begin{theorem}[soundness]
	$\theory{T} \vdash s = t$ implies $\theory{T}, \category{C} \Vdash s = t$.
	\qed
\end{theorem}

\subsection{Free Models}\label{subsec:free_model}

We describe a construction of a free model $\freefunctor{\theory{T}}X \in \modelcat{\theory{T}}(\ordMSet{M})$ of a graded theory $\theory{T}$ generated by a set $X$, which induces an adjunction between $\modelcat{\theory{T}}(\ordMSet{M})$ and $\Set$.
This adjunction, together with the $\category{M}$-action of Corollary~\ref{cor:MSet_left_action}, gives a graded monad as described in~\cite{DBLP:conf/fossacs/FujiiKM16}.

\begin{mydefinition}[free model $\freefunctor{\theory{T}}X$]
Let $\theory{T} = (\Sigma, E)$ be an $\category{M}$-graded theory.
We define a functor $\freefunctor{\theory{T}}X : \category{M} \to \Set$ by $\freefunctor{\theory{T}} X m \coloneqq \Terms{\Sigma}{m}{X} / {\sim_{m}}$ for each $m \in \category{M}$ and any $X \in \Set$ where $s \sim_{m} t$ is the equivalence relation defined by $\theory{T} \vdash s = t$ and $\freefunctor{\theory{T}} X w ([t]_{m}) \coloneqq [c_{w}(t)]_{m'}$ for any $w : m \to m'$ where $[t]_{m}$ is the equivalence class of $t \in \Terms{\Sigma}{m}{X}$.
For each $f \in \Sigma_{n, m'}$, let $|f|^{\freefunctor{\theory{T}}X} : (\freefunctor{\theory{T}} X)^n \to m' \act \freefunctor{\theory{T}}X$ be a mapping defined by
$|f|^{\freefunctor{\theory{T}}X}_m ([t_1]_{m}, \dots, [t_n]_{m}) = [f(t_1, \dots, t_n)]_{m' \mult m}$ for each $m \in \category{M}$.
We define a model of $\theory{T}$ by $\freefunctor{\theory{T}}X = (\freefunctor{\theory{T}}X, {|\cdot|}^{\freefunctor{\theory{T}}X})$.
\end{mydefinition}

The model $\freefunctor{\theory{T}}X$, together with the mapping $\eta_X : X \to \freefunctor{\theory{T}} X \identity$ defined by $x \mapsto [x]_{\identity}$, has the following universal property as a free model generated by $X$.
\begin{mylemma}
\label{lem:free_model_universal}
For any model $A$ in $\ordMSet{M}$ and any mapping $v : X \to A \identity$, there exists a unique homomorphism $\overline{v} : \freefunctor{\theory{T}}X \to A$ satisfying $\overline{v}_{\identity} \comp \eta_X = v$.
\qed
\end{mylemma}

\begin{mycorollary}
	Let $U : \modelcat{\theory{T}}(\ordMSet{M}) \to \Set$ be the forgetful functor defined by the evaluation at $\identity$, that is, $U A = A_{\identity}$ and $U \alpha = \alpha_{\identity}$.
	The free model functor $\freefunctor{\theory{T}} : \Set \to \modelcat{\theory{T}}(\ordMSet{M})$ is a left adjoint of $U$.
\qed
\end{mycorollary}

By considering the interpretation in the free model, we obtain the following completeness theorem.
\begin{mytheorem}[completeness]
	$\theory{T}, \ordMSet{M} \Vdash s = t$ implies $\theory{T} \vdash s = t$.
	\qed
\end{mytheorem}

Recall that $\modelcat{\theory{T}}(\ordMSet{M})$ has a left action (Corollary~\ref{cor:MSet_left_action}).
Therefore the above adjunction induces an $\category{M}$-graded monad as described in~\cite{DBLP:conf/fossacs/FujiiKM16}.

The relationship between $\modelcat{\theory{T}}(\ordMSet{M})$ and the Eilenberg--Moore construction is as follows.
In~\cite{DBLP:conf/fossacs/FujiiKM16}, the Eilenberg--Moore category $\category{C}^{\mathbf{T}}$ for any graded monad $\mathbf{T}$ on $\category{C}$ is introduced together with a left action ${\act} : \category{M} \times \category{C}^{\mathbf{T}} \to \category{C}^{\mathbf{T}}$.
If $\category{C} = \Set$ and $\mathbf{T}$ is the graded monad obtained from an $\category{M}$-graded theory $\theory{T}$, then the Eilenberg--Moore category $\Set^{\mathbf{T}}$ is essentially the same as $\modelcat{\theory{T}}(\ordMSet{M})$.
\begin{mytheorem}
	The comparison functor $K : \modelcat{\theory{T}}(\ordMSet{M}) \to \Set^{\mathbf{T}}$ (see~\cite{DBLP:conf/fossacs/FujiiKM16} for the definition) where $\theory{T}$ is an $\category{M}$-graded theory and $\mathbf{T}$ is the graded monad induced from the graded theory $\theory{T}$ is isomorphic.
	Moreover, $K$ preserves the $\category{M}$-action: ${\act} \comp (\monoid{M} \times K) = K \comp {\act}$.
	\qed
\end{mytheorem}

We define the category $\GradedSpec_{\category{M}}$ of graded algebraic theories as follows.
\begin{mydefinition}
Let $\theory{T} = (\Sigma, E)$ and $\theory{T}' = (\Sigma', E')$.
A morphism $\alpha : \theory{T} \to \theory{T}'$ between graded algebraic theories is a family of mappings $\alpha_{n, m} : \Sigma_{n, m} \to \freefunctor{\theory{T'}} n m$ from operations in $\Sigma$ to $\Sigma'$-terms such that the equations in $E$ are preserved by $\alpha$, i.e.\ for each $s, t \in \Terms{\Sigma}{m}{X}$, $(s, t) \in E$ implies $|s|^{(\freefunctor{\theory{T}'} X, \alpha)} = |t|^{(\freefunctor{\theory{T}'} X, \alpha)}$ where $(\freefunctor{\theory{T}'} X, \alpha)$ is a model of $\theory{T}$ induced by $\alpha$.
\end{mydefinition}

\begin{mydefinition}
	Given a morphism $\alpha : \theory{T} \to \theory{T}'$, let $\freefunctor{\alpha} : \freefunctor{\theory{T}} \to \freefunctor{\theory{T}'}$ be a natural transformation defined by $\freefunctor{\alpha}([t]) = |t|^{(\freefunctor{\theory{T}'} X, \alpha)}$ for each $t \in \Terms{\Sigma}{m}{X}$.
\end{mydefinition}

\begin{mydefinition}
	We write $\GradedSpec_M$ for the category of graded algebraic theories and morphisms between them.
	The identity morphisms are defined by $1_{\theory{T}}(f) = [f(x_1, \dots, x_n)]$ for each $f \in \Sigma_{n, m}$.
	The composition of $\alpha : \theory{T} \to \theory{T}'$ and $\beta : \theory{T}' \to \theory{T}''$ is defined by $\beta \comp \alpha (f) = \freefunctor{\beta}(\alpha(f))$.
\end{mydefinition}

\subsection{Examples}

\begin{myexample}[graded modules]
Let $\category{M} = (\nat, {+}, 0)$ where $\nat$ is regarded as a discrete category.
Given a graded ring $A = \bigoplus_{n \in \nat} A_n$, let $\Sigma$ be a set of operations which consists of the binary addition operation $+$ (arity: 2, grade: 0), the unary inverse operation $-$ (arity: 1, grade: 0), the identity element (nullary operation) $0$ (arity: 0, grade: 0) and the unary scalar multiplication operation $a \cdot ({-})$ (arity: 1, grade: $n$) for each $a \in A_n$.
Let $E$ be the equational axiom for modules.

A model $(F, |\cdot|)$ of the $\category{M}$-graded theory $(\Sigma, E)$ in $\ordMSet{M}$ consists of a set $F_n$ for each $n \in \nat$ and functions $|{+}|_n : (F_n)^2 \to F_n$, $|{-}|_n : F_n \to F_n$, $|0|_n \in F_n$ and $|a \cdot ({-})|_n : F_n \to F_{m + n}$ for each $n \in \nat$ and each $a \in A_m$, and these interpretations satisfy $E$.
Therefore models of $(\Sigma, E)$ in $\ordMSet{M}$ correspond one-to-one with graded modules.
\end{myexample}

\begin{myexample}[graded exception monad~{\cite[Example~3.4]{DBLP:conf/popl/Katsumata14}}]\label{ex:theory_graded_exception_monad}
{
\newcommand{\raisex}{\mathsf{raise}}
We give an algebraic presentation of the graded exception monad.

Let $\category{M}$ and $({*}, \eta, \mu)$ be a preordered monoid and the graded monad defined as follows.
Let $P^{+}(X)$ denote the set of nonempty subsets of $X$.
Let $\mathrm{Ex}$ be a set of exceptions and $\category{M} = ( (P^{+}(\mathrm{Ex} \cup \{ \mathrm{Ok} \}), {\subseteq}), \identity, {\mult})$ be a preordered monoid where $\identity = \{ \mathrm{Ok} \}$ and the multiplication ${\mult}$ is defined by $m \mult m' = (m \setminus \{ \mathrm{Ok} \}) \cup m'$ if $\mathrm{Ok} \in m$ and $m \mult m' = m$ otherwise (note that this is not commutative).
The graded exception monad $({*}, \eta, \mu)$ is the $\category{M}$-graded monad given as follows.%
\vspace{-2pt}
\begin{gather*}
m \ast X = \{ \mathrm{Er}(e) \mid e \in m \setminus \{ \mathrm{Ok} \} \} \cup \{ \mathrm{Ok}(x) \mid x \in X \land \mathrm{Ok} \in m \} \\
\eta_X(x) = \mathrm{Ok}(x) \qquad
\mu_{m_1, m_2, X}(\mathrm{Er}(e)) = \mathrm{Er}(e) \quad \mu_{m_1, m_2, X}(\mathrm{Ok}(x)) = x
\end{gather*}

The $\category{M}$-graded theory $\theory{T}^{\mathrm{ex}}$ for the graded exception monad is defined by $(\Sigma^{\mathrm{ex}}, \emptyset)$ where $\Sigma^{\mathrm{ex}}$ is the set that consists of an operation $\raisex_e$ (arity: 0, grade: $\{ e \}$) for each $e \in \mathrm{Ex}$.

The graded monad induced by $\theory{T}^{\mathrm{ex}}$ coincides with the graded exception monad.
Indeed, the free model functor $\freefunctor{\theory{T}^{\mathrm{ex}}}$ for $\theory{T}^{\mathrm{ex}}$ is given by $\freefunctor{\theory{T}^{\mathrm{ex}}} X m = m \mathrel{\ast} X$.
Here, the operations $\raisex_e$ are interpreted by $e \in \mathrm{Ex}$.
\vspace{-2pt}
\[ |\raisex_e|^{\freefunctor{\theory{T}^{\mathrm{ex}}}X}_m = \mathrm{Er}(e) \in \freefunctor{\theory{T}^{\mathrm{ex}}} X (\{ e \} \mult m) \]
}
\end{myexample}

\begin{myexample}[extending an ordinary monad to an $\category{M}$-graded monad]\label{ex:ordinary_monad_as_graded}
	We consider the problem of extending an $\category{M'}$-graded theory to an $\category{M}$-graded theory along a lax monoidal functor of type $\category{M'} \to \category{M}$, but here we restrict ourselves to the case of $\category{M'} = \mathbf{1}$ and the strict monoidal functor of type $\mathbf{1} \to \category{M}$.

Let $\category{M} = (\category{M}, \identity, {\mult})$ be an arbitrary small strict monoidal category.
Let $\theory{T} = (\Sigma, E)$ be a ($\mathbf{1}$-graded) theory and $(T, \eta^T, \mu^T)$ be the corresponding ordinary monad.
Let $\theory{T}^{\category{M}} = (\Sigma^{\category{M}}, E^{\category{M}})$ be the $\category{M}$-graded theory obtained when we regard each operation in $\theory{T}$ as an operation with grade $\identity \in \category{M}$, that is, $\Sigma^{\category{M}}_{n, m} \coloneqq \Sigma_n$ if $m = \identity$ and $\Sigma^{\category{M}}_{n, m} \coloneqq \emptyset$ otherwise, and $E^{\category{M}} \coloneqq E$.

The free model functor for $\theory{T}^{\category{M}}$ is $\freefunctor{\theory{T}^{\category{M}}} X = \freefunctor{\theory{T}}(\category{M}(\identity, {-}) \times X)$ where $\freefunctor{\theory{T}} : \Set \to \modelcat{\theory{T}}(\Set)$ is the free model functor for $\theory{T}$ as a $\mathbf{1}$-graded theory, and the interpretation of an operation $f \in \Sigma_n$ in $\freefunctor{\theory{T}^{\category{M}}} X$ is defined by the interpretation in the free models of $\theory{T}$.
\vspace{-2pt}
\[ |f|^{\freefunctor{\theory{T}^{\category{M}}} X}_m = |f|^{\freefunctor{\theory{T}}(\category{M}(\identity, m) \times X)} : \big( \freefunctor{\theory{T}}(\category{M}(\identity, m) \times X) \big)^n \to \freefunctor{\theory{T}}(\category{M}(\identity, m) \times X) \]
\vspace{-2pt}
Intuitively, this can be understood as follows.
Since all the operations are of grade $\identity$, coercions $c_w$ in a term can be moved to the innermost places where variables occur by repeatedly applying  $c_w(f(t_1, \dots, t_n)) = f(c_w(t_1), \dots, c_w(t_n))$ (see Definition~\ref{def:equational_logic}).
Therefore, we can consider terms of $\theory{T}^{\category{M}}$ as terms of $\theory{T}$ whose variables are of the form $c_w(x)$.

An $\category{M}$-graded monad $(\ast, \eta, \mu)$ obtained from $\theory{T}^{\category{M}}$ is as follows.
\vspace{-2pt}
\[ m \ast X = T(\category{M}(\identity, m) \times X) \qquad \eta = \eta^T(1_{\identity}, {-}) \qquad \mu = T({\mult} \times X) \comp \mu^T \comp T \mathrm{st} \]
\vspace{-2pt}
Here, ${\mult} : \category{M}(\identity, m_1) \times \category{M}(\identity, m_2) \to \category{M}(\identity, m_1 \mult m_2)$ is induced by $\mult : \category{M} \times \category{M} \to \category{M}$ and $\mathrm{st}_{X, Y} : X \times TY \to T(X \times Y)$ is the strength for $T$.
\end{myexample}

\section{Graded Lawvere Theories}\label{sec:graded_lawvere_theory}

We present a categorical formulation of graded algebraic theories of \S\ref{sec:graded_algebraic_theory} in a similar fashion to ordinary Lawvere theories.

For ordinary (single-sorted) finitary algebraic theories, a \emph{Lawvere theory} is defined as a small category $\category{L}$ with finite products together with a strict finite-product preserving identity-on-objects functor $J : \aleph_0^{\op} \to \category{L}$ where $\aleph_0$ is the full subcategory of $\Set$ on natural numbers.
Intuitively, morphisms in the Lawvere theory $\category{L}$ are terms of the corresponding algebraic theory, and objects of $\category{L}$, which are exactly the objects in $\obj \aleph_0$, are arities.

According to the above intuition, it is expected that a graded Lawvere theory is also defined as a category whose objects are natural numbers and morphisms are graded terms.
However, since terms in a graded algebraic theory are stratified by a monoidal category $\category{M}$, mere sets are insufficient to express hom-objects of graded Lawvere theories.
Instead, we take hom-objects from the functor category $\ordMSet{M}$ and define graded Lawvere theories using $\ordMSet{M}$-categories where $\ordMSet{M}$ is equipped with the Day convolution monoidal structure.
Specifically, $\aleph_0$ (in ordinary Lawvere theories) is replaced with an $\ordMSet{M}$-category $\initialGLawop{\category{M}}$, $\category{L}$ with an $\ordMSet{M}$-category, and ``finite products'' with ``$\initialGLaw{\category{M}}$-cotensors''.

So, we first provide an enriched category $\initialGLawop{\category{M}}$ that we use as arities.
Since we do not assume that $\category{M}$ is symmetric, $\initialGLawop{\category{M}}$ is defined to be an $\twist{\ordMSet{M}}$-category so that the opposite category $\initialGLaw{\category{M}}$ is an $\ordMSet{M}$-category.
Let $\twist{\MSet{M}}$ be an $\twist{\ordMSet{M}}$-category induced by the closed structure of $\twist{\ordMSet{M}}$.
That is, hom-objects of $\twist{\MSet{M}}$ are given by $\twist{\MSet{M}}(G, H) m = \ordMSet{M}(G, H({-} \mult m))$.

\begin{mydefinition}
An $\twist{\ordMSet{M}}$-category $\initialGLawop{\category{M}}$ is defined by the full sub-$\twist{\ordMSet{M}}$-category of $\twist{\MSet{M}}$ whose set of objects is given by $\obj \initialGLawop{\category{M}} = \{ n \cdot y(\identity) \mid n \in \nat \} \subseteq \obj \twist{\MSet{M}}$ where $\nat$ is the set of natural numbers and $n \cdot y(\identity)$ is the $n$-fold coproduct of $y(\identity)$.
We sometimes identify $\obj \initialGLawop{\category{M}}$ with $\nat$ via the mapping $n \mapsto \underline{n} \coloneqq n \cdot y(\identity)$.
\end{mydefinition}

\begin{mylemma}
The $\ordMSet{M}$-category $\initialGLaw{\category{M}}$ has $\initialGLaw{\category{M}}$-cotensors, which are given by $\underline{n} \cotensor \underline{n'} = \underline{n \cdot n'}$ for each $n$ and $n'$.
\qed
\end{mylemma}
\begin{proof}
	A cotensor $(n \cdot y(\identity)) \cotensor (n' \cdot y(\identity))$ is a tensor $(n \cdot y(\identity)) \mathrel{\twist{\mult}} (n' \cdot y(\identity))$ in $\twist{\MSet{M}}$.
	Since $\twist{\mult}$ is biclosed, $\twist{\mult}$ preserves colimits in both arguments.
	Therefore, $(n \cdot y(\identity)) \mathrel{\twist{\mult}} (n' \cdot y(\identity)) \cong (n \cdot n') \cdot y(\identity)$.
	\qed
\end{proof}

$\initialGLaw{\category{M}}$-cotensors (i.e.\ $n \cdot y(\identity) \cotensor C$) behave like an enriched counterpart of finite powers $({-})^n$.
We show that $\initialGLaw{\category{M}}$-cotensors in a general $\ordMSet{M}$-category $\category{A}$ are characterized by projections satisfying a universal property.
Given a unit morphism $\nu : \underline{n} \to \category{A}(\underline{n} \cotensor C, C)$ of the cotensor $\underline{n} \cotensor C$, an $\ordMSet{M}$-natural transformation $\overline{\nu} : \category{A}(B, \underline{n} \cotensor C) \to \rclosed{\underline{n}}{\category{A}(B, C)}$ is given by $f \mapsto (x \mapsto \nu(x) \comp f)$.
The condition that $\overline{\nu}$ is isomorphic can be rephrased as follows.

\begin{mylemma}\label{lem:N_M_cotensor}
An $\ordMSet{M}$-category $\category{A}$ has $\initialGLaw{\category{M}}$-cotensors if and only if for any $n \in \nat$ and $C \in \obj{\category{A}}$, there exist an object $\underline{n} \cotensor C \in \obj{\category{A}}$ and $(\pi_1, \dots, \pi_n) \in (\category{A}(\underline{n} \cotensor C, C) I)^n$ such that the following condition holds: for each $m$, the function $f \mapsto (\pi_1 \comp f, \dots, \pi_n \comp f)$ of type $\category{A}(B, \underline{n} \cotensor C) m \to (\category{A}(B, C) m)^n$
is bijective.

An $\ordMSet{M}$-functor $F : \category{A} \to \category{B}$ preserves $\initialGLaw{\category{M}}$-cotensors if and only if $(F_{\underline{n} \cotensor C, C, I} \comp \pi_1, \dots, F_{\underline{n} \cotensor C, C, I} \comp \pi_n) \in (\category{B}(F(\underline{n} \cotensor C), F C) I)^n$ satisfies the same condition for each $n$ and $C$.
\end{mylemma}
\begin{proof}
The essence of the proof is that the unit morphism $\nu : n \cdot y(\identity) \to \category{A}(\underline{n} \cotensor C, C)$ corresponds to elements $\pi_1, \dots, \pi_n \in \category{A}(\underline{n} \cotensor C, C) \identity$ by $\ordMSet{M} (n \cdot y(\identity), \category{A}(\underline{n} \cotensor C, C)) \cong \ordMSet{M} (y(\identity), \category{A}(\underline{n} \cotensor C, C))^n \cong \big( \category{A}(\underline{n} \cotensor C, C) \identity \big)^n$.
The $\ordMSet{M}$-natural transformation $\overline{\nu}$ is isomorphic if and only if each component $\overline{\nu}_{m} : \category{A}(B, \underline{n} \cotensor C)m \to \rclosed{\underline{n}}{\category{A}(B, C)} m$ of $\overline{\nu}$ is isomorphic, which is moreover equivalent to the condition that $f \mapsto (\pi_1 \comp f, \dots, \pi_n \comp f) : \category{A}(B, \underline{n} \cotensor C)m \to (\category{A}(B, C))^n$ is isomorphic since we have $\rclosed{\underline{n}}{\category{A}(B, C)} m \cong (\category{A}(B, C) m)^n$.

The latter part of the lemma follows from the former part.
\qed
\end{proof}
If $(\pi_1, \dots, \pi_n) \in (\category{A}(\underline{n} \cotensor C, C) \identity)^n$ satisfies the condition in Lemma~\ref{lem:N_M_cotensor}, we call the element $\pi_i \in \category{A}(\underline{n} \cotensor C, C) \identity$ the \emph{$i$-th projection} of $\underline{n} \cotensor C$.
Note that the choice of projections is not necessarily unique.
However, when we say that $\category{A}$ is an $\ordMSet{M}$-category with $\initialGLaw{\category{M}}$-cotensors, we implicitly assume that there are a chosen cotensor $\underline{n} \cotensor C$ and chosen projections $(\pi_1, \dots, \pi_n) \in (\category{A}(\underline{n} \cotensor C, C) \identity)^n$ for each $\underline{n} \in \obj \initialGLaw{\category{M}}$ and $C \in \obj \category{A}$.
We also assume that $\underline{1} \cotensor X = X$ without loss of generality.
Given $n$-tuple $(f_1, \dots, f_n)$ of elements in $\category{A}(B, C) m$, we denote by $\langle f_1, \dots, f_n \rangle$ an element in $\category{A}(B, \underline{n} \cotensor C) m$ obtained by the inverse of $f \mapsto (\pi_1 \comp f, \dots, \pi_n \comp f)$ and call this a tupling.
Tuplings and projections for $\initialGLaw{\category{M}}$-cotensors behave like those for finite products.

The following proposition claims that $\initialGLaw{\category{M}}$ is a free $\ordMSet{M}$-category with chosen $\initialGLaw{\category{M}}$-cotensors generated by one object.
\begin{myproposition}\label{prop:functor_from_initialGLaw}
Let $\category{A}$ be an $\ordMSet{M}$-category with $\initialGLaw{\category{M}}$-cotensors and $C$ be an object in $\category{A}$.
Then there exists a unique $\initialGLaw{\category{M}}$-cotensor preserving $\ordMSet{M}$-functor $F : \initialGLaw{\category{M}} \to \category{A}$ such that $F n = \underline{n} \cotensor C$ and $F \pi_i = \pi_i$.
\qed
\end{myproposition}

We define $\category{M}$-graded Lawvere theories in a similar fashion to enriched Lawvere theories.
\begin{mydefinition}\label{def:graded_lawvere_theory}
	An $\category{M}$-graded Lawvere theory is a tuple $(\category{L}, J)$ where $\category{L}$ is an $\ordMSet{M}$-category with $\initialGLaw{\category{M}}$-cotensors and $J : \initialGLaw{\category{M}} \to \category{L}$ is an identity-on-objects $\initialGLaw{\category{M}}$-cotensor preserving $\ordMSet{M}$-functor.
A morphism $F : (\category{L}, J) \to (\category{L'}, J')$ between two graded Lawvere theories is an $\ordMSet{M}$-functor $F : \category{L} \to \category{L'}$ such that $FJ = J'$.
We denote the category of graded Lawvere theories and morphisms between them by $\GradedLawvere_{\category{M}}$.
\end{mydefinition}

By Proposition~\ref{prop:functor_from_initialGLaw}, the existence of the above $J : \initialGLaw{\category{M}} \to \category{L}$ is equivalent to requiring that $\obj \category{L} = \nat$ and projections in $\category{L}$ are chosen in some way.
So, we sometimes leave $J$ implicit and just write $\category{L} \in \GradedLawvere_{\category{M}}$ for $(\category{L}, J) \in \GradedLawvere_{\category{M}}$.

\begin{mydefinition}
	A \emph{model} of graded Lawvere theory $\category{L}$ in an $\ordMSet{M}$-category $\category{A}$ with $\initialGLaw{\category{M}}$-cotensor is an $\initialGLaw{\category{M}}$-cotensor preserving $\ordMSet{M}$-functor of type $\category{L} \to \category{A}$.
A morphism $\alpha : F \to G$ between two models $F, G$ of graded Lawvere theory $\category{L}$ is an $\ordMSet{M}$-natural transformation.
Let $\mathrm{Mod}(\category{L}, \category{A})$ be the category of models of graded Lawvere theory $\category{L}$ in the $\ordMSet{M}$-category $\category{A}$.
\end{mydefinition}

In \S\ref{sec:graded_algebraic_theory}, we use a category $\category{C}$ satisfying $\category{M}$-model condition to define a model of graded algebraic theory.
Actually, $\category{M}$-model condition is sufficient to give an $\ordMSet{M}$-category with $\initialGLaw{\category{M}}$-cotensors.
\begin{mylemma}\label{lem:enrichment_by_action_has_contensors}
	If $\category{C}$ satisfies $\category{M}$-model condition, then the $\ordMSet{M}$-category $\KleisliLike{\category{C}}{{T}}^{\op}$ defined in Definition~\ref{def:enrichment_by_action} has $\initialGLaw{\category{M}}$-cotensors.
\end{mylemma}
\begin{proof}
	For any $X \in \KleisliLike{\category{C}}{{T}}^{\op}$ and $n$, the cotensor $\underline{n} \cotensor X$ is given by finite power $X^n$, and the $i$-th projection is given by $\eta^{\act} \comp \pi_i \in \KleisliLike{\category{C}}{{T}}^{\op} \identity$ where $\pi_i : X^n \to X$ is the $i$-th projection of the finite power $X^n$.
	The rest of the proof is routine.
	\qed
\end{proof}

If we apply Lemma~\ref{lem:enrichment_by_action_has_contensors} to $\ordMSet{M}$ equipped with the $\twist{\category{M}}$-action in Example~\ref{example:functor_cat_right_action} (here denoted by $T$), then $\KleisliLike{(\ordMSet{M})}{T}^{\op}$ coincides with $\MSet{M}$ (i.e.\ the $\ordMSet{M}$-category obtained by the closed structure of $\ordMSet{M}$).

\section{Equivalence}\label{sec:equivalence}

We have shown three graded notions: graded algebraic theories, graded Lawvere theories and finitary graded monads, which give rise to categories $\GradedSpec_{\category{M}}$, $\GradedLawvere_{\category{M}}$ and $\GradedMonad_{\category{M}}$, respectively.
This section is about the equivalence of these three notions.
We give only a sketch of the proof of the equivalence, and the details are deferred to \referappendix{sec:appendix:proof_equivalence}{A}.

\subsection{Graded Algebraic Theories and Graded Lawvere Theories}
We prove that the category of graded algebraic theories $\GradedSpec_{\category{M}}$ and the category of graded Lawvere theories $\GradedLawvere_{\category{M}}$ are equivalent by showing the existence of an adjoint equivalence $\theoryfunctor{} \vdash U : \GradedLawvere_{\category{M}} \to \GradedSpec_{\category{M}}$.

Let $\category{M}$ be a small strict monoidal category and $\theory{T} = (\Sigma, E)$ be an $\category{M}$-graded algebraic theory.
We define $\theoryfunctor \theory{T}$ (the object part of $\theoryfunctor$) as an $\category{M}$-graded Lawvere theory whose morphisms are terms of $\theory{T}$ modulo equational axioms.

\begin{mydefinition}
	An $\ordMSet{M}$-category $\theoryfunctor \theory{T}$ is defined by $\mathrm{ob}(\theoryfunctor \theory{T}) \coloneqq \nat$ and $(\theoryfunctor \theory{T})(n, n') m \coloneqq (\freefunctor{\theory{T}} n  m)^{n'}$ with composition defined by substitution.
\end{mydefinition}

It is easy to show that $\theoryfunctor{\theory{T}}$ has $\initialGLaw{\category{M}}$-cotensors (by Lemma~\ref{lem:N_M_cotensor}).
Therefore, $\theoryfunctor{}$ is a mapping from an object in $\GradedSpec_{\category{M}}$ to an object in $\GradedLawvere_{\category{M}}$.

We define a functor $U : \GradedLawvere_M \to \GradedSpec_M$ by taking all the morphism $f \in L(n, 1) m$ in $L \in \GradedLawvere_M$ as operations and all the equations that hold in $L$ as equational axioms.

\begin{mydefinition}
A functor $U : \GradedLawvere_M \to \GradedSpec_M$ is defined as follows.
\begin{itemize}
	\item For each $\category{L} \in \obj \GradedLawvere_M$, $U \category{L} = (\Sigma, E)$ where $\Sigma_{n, m} = \category{L}(n, 1)m$, $E = \{ (s, t) \mid |s|^\category{L} = |t|^\category{L} \}$ and $|\cdot|^\category{L} : T^{\Sigma}_m(n) \to \category{L}(n, 1) m$ is an interpretation of terms defined in the same way as Definition~\ref{def:term_interpretation}.
\item Given $G : \category{L} \to \category{L'}$, let $UG : U\category{L} \to U\category{L'}$ be a functor defined by
$UG (f) = [G(f)(x_1, \dots, x_n)]$ for each $f \in \category{L}(n, 1) m$.
\end{itemize}
\end{mydefinition}

Then, $\theoryfunctor{\theory{T}}$ has the following universal property as a left adjoint of $U$.
\begin{mylemma}\label{lem:ThT_universal}
For each $\theory{T}$, let $\eta_{\theory{T}} : \theory{T} \to U \theoryfunctor \theory{T}$ be a family of functions $\eta_{\theory{T}, n, m} : \Sigma_{n, m} \to \freefunctor{U \theoryfunctor \theory{T}} n  m$ defined by $\eta_{\theory{T}, n, m}(f) = [[f(x_1, \dots, x_n)](x_1, \dots, x_n)]$.
For any $\alpha : \theory{T} \to U\category{L}$, there exists a unique morphism $\overline{\alpha} : \theoryfunctor \theory{T} \to \category{L}$ such that $\alpha = U \overline{\alpha} \comp \eta_{\theory{T}}$.
\qed
\end{mylemma}

Moreover, the unit and the counit of $\theoryfunctor \dashv U$ are isomorphic.
Therefore:
\begin{mytheorem}\label{thm:GS_GLaw_equiv}
Two categories $\GradedSpec_{\category{M}}$ and $\GradedLawvere_{\category{M}}$ are equivalent.
\qed
\end{mytheorem}

We can also prove the equivalence of the categories of models.
\begin{mylemma}
	If $\category{C}$ is a category satisfying $\category{M}$-model condition, then $\modelcat{\theory{T}}(\category{C})$ is equivalent to $\mathrm{Mod}(\theoryfunctor \theory{T}, \KleisliLike{\category{C}}{T})$ where $T$ is the $\twist{\category{M}}$-action on $\category{C}$.
	\qed
\end{mylemma}

\subsection{Graded Lawvere theories and Finitary Graded Monads}
We prove that the category of graded Lawvere theories $\GradedLawvere_{\category{M}}$ and the category of finitary graded monads $\GradedMonad_{\category{M}}$ are equivalent.
Given a graded Lawvere theory, a finitary graded monad is obtained as a coend that represents the set of terms.
On the other hand, given a finitary graded monad, a graded Lawvere theory is obtained from taking the full sub-$\ordMSet{M}$-category on arities $\obj (\initialGLaw{\category{M}})$ of the opposite category of the Kleisli(-like) category in Definition~\ref{def:enrichment_by_action}.
These constructions give rise to an equivalence of categories.

An $\category{M}$-graded Lawvere theory yields a finitary graded monad by letting $m \ast X$ be the set of terms of grade $m$ whose variables range over $X$.
\begin{mydefinition}
	Let $\category{L}$ be an $\category{M}$-graded Lawvere theory.
	We define $T_{\category{L}} = (\ast, \eta, \mu)$ by a (finitary) $\category{M}$-graded monad whose functor part is given as follows.
	\vspace{-3pt}
	\[ m \ast X \coloneqq \int^{n \in \aleph_0} \category{L}(\underline{n}, \underline{1})m \times X^n \]
	\vspace{-3pt}
	Note that $\category{L}(\underline{{-}}, \underline{1}) : \aleph_0 \to \ordMSet{M}$ is a $\Set$-functor here.
\end{mydefinition}

Given a graded monad, a graded Lawvere theory is obtained as follows.

\begin{mydefinition}
Let $T = (\ast, \eta, \mu)$ be an $\category{M}$-graded monad on $\Set$.
Let $\category{L}_T$ be the full sub-$\ordMSet{M}$-category of $(\KleisliLike{\Set}{T})^{\op}$ with $\obj(\category{L}_T) = \nat$.
\end{mydefinition}

Since $\category{L}_T$ has $\category{N}_M$-cotensors $\underline{n} \cotensor 1 = n$ whose projections are given by $\pi_i = ({*} \mapsto \eta(i)) \in \Set(1, \identity \ast n)$, $\category{L}_T$ is a graded Lawvere theory.

Given a morphism $\alpha : T \to T'$ in $\GradedMonad_{\category{M}}$, we define $\category{L}_{\alpha} : \category{L}_T \to \category{L}_{T'}$ by $(\category{L}_{\alpha})_{n, n', m} = \Set(n', \alpha_{n, m}) : \category{L}_T(n, n') m \to \category{L}_{T'}(n, n') m$.
It is easy to prove that $\category{L}_{\alpha}$ is a morphism in $\GradedLawvere_{\category{M}}$ and $\category{L}_{({-})} : \GradedMonad_{\category{M}} \to \GradedLawvere_{\category{M}}$ is a functor.

\begin{mytheorem}\label{thm:GLaw_GMnd_equiv}
	Two categories $\GradedLawvere_{\category{M}}$ and $\GradedMonad_{\category{M}}$ are equivalent.
\end{mytheorem}
\begin{proof}
	$\category{L}_{({-})}$ is an essentially surjective fully faithful functor.
	\qed
\end{proof}

\section{Combining Effects}\label{sec:combining_effects}
Under the correspondence to algebraic theories, combinations of computational effects can be understood as combinations of algebraic theories.
In particular, sums and tensor products are well-known constructions~\cite{DBLP:journals/tcs/HylandPP06}.
In this section, we show that these constructions can be adapted to graded algebraic theories.
By the equivalence $\GradedMonad_{\category{M}} \equivcat \GradedLawvere_{\category{M}} \equivcat \GradedSpec_{\category{M}}$ in \S\ref{sec:equivalence}, constructions like sums and tensor products in one of these categories induce those in the other two categories.
So, we choose $\GradedSpec_{\category{M}}$ and describe sums as colimits in $\GradedSpec_{\category{M}}$ and tensor products as a mapping $\GradedSpec_{\category{M}_1} \times \GradedSpec_{\category{M}_2} \to \GradedSpec_{\category{M_1} \times \category{M}_2}$.

\subsection{Sums}
We prove that $\GradedSpec_{\category{M}}$ has small colimits.

\begin{mylemma}
The category $\GradedSpec_M$ has small coproducts.
\end{mylemma}
\begin{proof}
	Given a family $\{ (\Sigma^{(i)}, E^{(i)}) \}_{i \in I}$ of objects in $\GradedSpec_{\category{M}}$, the coproduct is obtained by the disjoint union of operations and equations: $\coprod_{i \in I} (\Sigma^{(i)}, E^{(i)}) = \left( \bigcup_{i \in I} \Sigma^{(i)}, \bigcup_{i \in I} E^{(i)} \right)$.
	\qed
\end{proof}

\begin{mylemma}
The category $\GradedSpec_M$ has coequalizers.
\end{mylemma}
\begin{proof}
	Let $\theory{T} = (\Sigma, E)$ and $\theory{T}' = (\Sigma', E')$ be graded algebraic theories and $\alpha, \beta : \theory{T} \to \theory{T}'$ be a morphism.
	The coequalizer $\theory{T}''$ of $\alpha$ and $\beta$ is given by adding the set of equations induced by $\alpha$ and $\beta$ to $\theory{T}'$, that is, $\theory{T}'' \coloneqq (\Sigma', E' \cup E'')$ where $E'' = \{ (s, t) \mid \exists f \in \Sigma, \alpha(f) = [s] \land \beta(f) = [t] \}$.
	\qed
\end{proof}

Since a category has all small colimits if and only if it has all small coproducts and coequalizers, we obtain the following corollary.
\begin{mycorollary}
Three equivalent categories $\GradedSpec_{\category{M}}$, $\GradedMonad_{\category{M}}$ and $\GradedLawvere_{\category{M}}$ are cocomplete.
\qed
\end{mycorollary}

\begin{myexample}
{
\newcommand{\raisex}{\mathsf{raise}}
It is known that the sum of an ordinary monad $T$ and the exception monad $({-}) + \mathrm{Ex}$ (where $\mathrm{Ex}$ is a set of exceptions) is given by $T(({-}) + \mathrm{Ex})$ \cite[Corollary 3]{DBLP:journals/tcs/HylandPP06}.
We show that a similar result holds for the graded exception monad.

Let $\theory{T}^{\mathrm{ex}}$ be the theory in Example~\ref{ex:theory_graded_exception_monad} and $\category{M}$ be the preordered monoid used there.
We denote $(\ast^{\mathrm{ex}}, \eta^{\mathrm{ex}}, \mu^{\mathrm{ex}})$ for the graded exception monad.
Let $\theory{T} = (\Sigma, E)$ be a ($\mathbf{1}$-graded) theory and $(T, \eta^T, \mu^T)$ be the corresponding ordinary monad.
Let $\theory{T}^{\category{M}} = (\Sigma^{\category{M}}, E^{\category{M}})$ be the $\category{M}$-graded theory obtained from $\theory{T}$ as in Example~\ref{ex:ordinary_monad_as_graded}.
We consider a graded monad obtained as the sum of $\theory{T}^{\mathrm{ex}}$ and $\theory{T}^{\category{M}}$.

A free model functor $F$ for $\theory{T}^{\mathrm{ex}} + \theory{T}^{\category{M}}$ is given by $F X m = T(m \ast^{\mathrm{ex}} X)$.
For each $n$-ary operation $f$ in $\theory{T}$, $|f|^{FX}_m : (T(m \ast^{\mathrm{ex}} X))^n \to T(m \ast^{\mathrm{ex}} X)$ is induced by free models of $\theory{T}$, and for each $e \in \mathrm{Ex}$, $|\raisex_e|^{FX}_m : 1 \to T(\{ e \} \ast^{\mathrm{ex}} X)$ is defined by $\eta^T_{\{e\} \ast^{\mathrm{ex}} X}(e) \in T(\{ e \} \ast^{\mathrm{ex}} X)$.
It is easy to see that $FX$ defined above is indeed a model of $\theory{T}^{\mathrm{ex}} + \theory{T}^{\category{M}}$.
Therefore, we obtain a graded monad $m * X = T(m \ast^{\mathrm{ex}} X)$.

}
\end{myexample}

\subsection{Tensor Products}\label{subsec:tensor_products}
The tensor product of two ordinary algebraic theories $(\Sigma, E)$ and $(\Sigma', E')$ is constructed as $(\Sigma \cup \Sigma', E \cup E' \cup E_{\tensor})$ where $E_{\tensor}$ consists of
$f(\lambda i. g(\lambda j. x_{i j})) = g(\lambda j. f(\lambda i. x_{i j}))$
for each $f \in \Sigma$ and $g \in \Sigma'$.
However, when we extend tensor products to graded algebraic theories, the grades of the both sides are not necessarily equal.
If the grade of $f$ is $m$ and the grade of $g$ is $m'$, then the grades of $f(\lambda i. g(\lambda j. x_{i j}))$ and $g(\lambda j. f(\lambda i. x_{i j}))$ are $m \mult m'$ and $m' \mult m$, respectively.
Therefore, we have to somehow guarantee that the grade of $f \in \Sigma$ and the grade of $g \in \Sigma'$ commute.
We solve this problem by taking the product of monoidal categories.
That is, we define the tensor product of an $\category{M}_1$-graded algebraic theory and an $\category{M}_2$-graded algebraic theory as an $\category{M}_1 \times \category{M}_2$-graded algebraic theory.

Before defining tensor products, we consider extending an $\category{M}$-graded theory to $\category{M'}$-graded theory along a lax monoidal functor $G = (G, \eta^G, \mu^G) : \category{M} \to \category{M'}$.
Given an $\category{M}$-graded theory $\theory{T} = (\Sigma, E)$, we define the $\category{M'}$-graded theory $G_{*}\theory{T} = (G_{*}\Sigma, G_{*}E)$ by
$(G_{*} \Sigma)_{n, m'} \coloneqq \{ f \in \Sigma_{n, m} \mid G m = m' \}$ and $G_{*} E \coloneqq \{ G_{*}(s) = G_{*}(t) \mid (s = t) \in E \}$
where for each term $t$ of $\theory{T}$ (with grade $m$), $G_{*}(t)$ is the term of $G_{*} \theory{T}$ (with grade $Gm$) defined inductively as follows:
if $x$ is a variable, then $G_{*}(x) \coloneqq c_{\eta^G}(x)$;
for each $w : m \to m'$ and term $t$, $G_{*}(c_w(t)) \coloneqq c_{Gw}(G_{*}(t))$;
for each $f \in \Sigma_{n, m}$ and terms $t_1, \dots, t_n$ with grade $m'$, $G_{*}(f(t_1, \dots, t_n)) \coloneqq c_{\mu^G_{m, m'}}(f(G_{*}(t_1), \dots, G_{*}(t_n)))$.

The tensor product of $\theory{T}_1 \,{\in}\, \GradedSpec_{\category{M}_1}$ and $\theory{T}_2 \,{\in}\, \GradedSpec_{\category{M}_2}$ is defined by first extending $\theory{T}_1$ and $\theory{T}_2$ to $\category{M}_1 {\times} \category{M}_2$-graded theories and then adding commutation equations.

\begin{mydefinition}[tensor product]
Let $\theory{T}_1 = (\Sigma, E) \in \GradedSpec_{\category{M}_1}$ and $\theory{T}_2 = (\Sigma', E') \allowbreak \in \GradedSpec_{\category{M}_2}$.
The \emph{tensor product} $\theory{T}_1 \otimes \theory{T}_2$ is defined by $(K_{*} \Sigma \cup K'_{*} \Sigma', K_{*} E \cup K'_{*} E' \cup E_{\theory{T}_1 \otimes \theory{T}_2}) \in \GradedSpec_{\category{M}_1 \times \category{M}_2}$ where $K : \category{M}_1 \to \category{M}_1 \times \category{M}_2$ and $K' : \category{M}_2 \to \category{M}_1 \times \category{M}_2$ are lax monoidal functors defined by $K m_1 \coloneqq (m_1, \identity_2)$ and $K' m_2 \coloneqq (\identity_1, m_2)$, and
{
\setlength{\abovedisplayskip}{3pt}
\setlength{\belowdisplayskip}{3pt}
\[ E_{\theory{T}_1 \otimes \theory{T}_2} \coloneqq \{ f(\lambda i. g(\lambda j. x_{ij})) = g(\lambda j. f(\lambda i. x_{ij})) \mid f \in (K_{*} \Sigma)_{n, m}, g \in (K'_{*} \Sigma')_{n', m'} \}. \]
}
\end{mydefinition}
That is, if $f$ is an operation in $\theory{T}_1$ with grade $m_1 \in \category{M}_1$, then $\theory{T}_1 \tensor \theory{T}_2$ has the operation $f$ with grade $(m_1, \identity_2) \in \category{M}_1 \times \category{M}_2$ and similarly for operations in $\theory{T}_2$.

The tensor products satisfy the following fundamental property.
\begin{myproposition}
	Let $\category{C}$ be a category satisfying $\category{M}_1 \times \category{M}_2$-model condition.
	Let $\theory{T}_i$ be an $\category{M}_i$-graded algebraic theory for $i = 1, 2$.
	Then we have an isomorphism $\modelcat{\theory{T}_1}(\modelcat{\theory{T}_2}(\category{C})) \cong \modelcat{\theory{T}_1 \tensor \theory{T}_2}(\category{C})$.
\end{myproposition}
\begin{proof}
	Let $((A, |\cdot|'), |\cdot|) \in \modelcat{\theory{T}_1}(\modelcat{\theory{T}_2}(\category{C}))$ be a model.
	For each operation $f$ in $\theory{T}_1$, $|f| : (A, |\cdot|')^n \to m \act (A, |\cdot|')$ is a homomorphism.
	This condition is equivalent to satisfying the equations in $E_{\theory{T}_1 \tensor \theory{T}_2}$.
	\qed
\end{proof}

\begin{myexample}
{
\newcommand{\lookup}{\mathsf{lookup}}
\newcommand{\update}{\mathsf{update}}

We exemplify the tensor product by showing a graded version of~\cite[Corollary~6]{DBLP:journals/tcs/HylandPP06}, which claims that the $L$-fold tensor product of the side-effects theory in~\cite{DBLP:conf/fossacs/PlotkinP02} with one location is the side-effects theory with $L$ locations.

First, we consider the situation where there is only one memory cell whose value ranges over a finite set $V$.
Let $\mathbf{2}$ the preordered monoid (join-semilattice) $(\{ \bot, \top \}, {\le}, {\lor}, \bot)$ where ${\leq}$ is the preorder defined by $\bot \leq \top$.
Intuitively, $\bot$ represents pure computations, and $\top$ represents (possibly) stateful computations.
Let $\theory{T}_{\mathrm{st}}$ be a $\mathbf{2}$-graded theory of two types of operations
$\lookup$ (arity: $V$, grade: $\top$) and
$\update_v$ (arity: $1$, grade: $\top$) for each $v \in V$
and the four equations in~\cite{DBLP:conf/fossacs/PlotkinP02} for the interaction of $\lookup$ and $\update$.
Note that we have to insert coercion to arrange the grade of the equation $\lookup(\lambda v \in V. \update_v(x)) = c_{\bot \le \top}(x)$.

The graded monad $(\ast, \eta, \mu)$ induced by $\theory{T}_{\mathrm{st}}$ is as follows.
{
\setlength{\abovedisplayskip}{3pt}
\setlength{\belowdisplayskip}{3pt}
\[ \bot \ast X = X \qquad \top \ast X = (V \times X)^V \qquad ((\bot \le \top) \ast X)(x) = \lambda v. (v, x) \]
}
The middle equation can be explained as follows: any term with grade $\top$ can be presented by a canonical form $t_f \coloneqq \lookup(\lambda v. \update_{f_V(v)}(f_X(v)))$ where $f = \langle f_V, f_X \rangle : V \to V \times X$ is a function, and therefore, the mapping $f \mapsto t_f$ gives a bijection between $(V \times X)^V$ and $\top * X = \Terms{\Sigma}{\top}{X}/{\sim}$.

The $L$-fold tensor product of $\theory{T}_{\mathrm{st}}$, which we denote by $\theory{T}_{\mathrm{st}}^{{\tensor} L}$, is a $\mathbf{2}^L$-graded theory where $\mathbf{2}^L = (2^L, \subseteq, {\cup}, \emptyset)$ is the join-semilattice of subsets of $L$.
Specifically, $\theory{T}_{\mathrm{st}}^{{\tensor} L}$ consists of operations $\lookup_l$ and $\update_{l, v}$ with grade $\{ l \}$ for each $l \in L$ and $v \in V$ with additional three commutation equations in~\cite{DBLP:conf/fossacs/PlotkinP02}.
The induced graded monad is
$L' \ast^{{\tensor} L} X = \{ f : V^L \to (V^L \times X) \mid \mathrm{read}(L', f) \land \mathrm{write}(L', f) \}$
where $L' \subseteq L$, and $\mathrm{read}(L', f)$ and $\mathrm{write}(L', f)$ assert that $f$ depends only on values at locations in $L'$ and does not change values at locations outside $L'$.
That is, $L' \ast^{{\tensor} L} X$ represents computations that touch only memory locations in $L'$.%
\vspace{-3pt}
\begin{align*}
\mathrm{read}(L', f) &\quad\coloneqq\quad \forall \sigma, \sigma' \in V^n, (\forall l \in L', \sigma(l) = \sigma'(l)) \implies f(\sigma) = f(\sigma') \\
\mathrm{write}(L', f) &\quad\coloneqq\quad \forall \sigma, \sigma' \in V^n, x \in X, (\sigma', x) = f(\sigma) \implies \forall l \notin L', \sigma(l) = \sigma'(l)
\end{align*}
}
\end{myexample}

\section{Related Work}\label{sec:related}

\paragraph{Algebraic theories for graded monads.}
Graded monads are introduced in~\cite{smirnov2008graded}, and notions of graded theory and graded Eilenberg--Moore algebra appear in~\cite{DBLP:conf/calco/MiliusPS15,DBLP:conf/concur/DorschMS19} for coalgebraic treatment of trace semantics.
However, these work only deal with $\nat$-graded monads where $\nat$ is regarded as a discrete monoidal category, while we deal with general monoidal categories.
The Kleisli construction and the Eilenberg--Moore construction for graded monads are presented in~\cite{DBLP:conf/fossacs/FujiiKM16} by adapting the 2-categorical argument on resolutions of monads~\cite{STREET1972149}.

Algebraic operations for graded monads are introduced in~\cite{DBLP:conf/popl/Katsumata14} and classified into two types, which are different in how to integrate the grades of subterms.
One is operations that take terms with the same grade, and these are what we treated in this paper.
The other is operations that take terms with different grades: the grade of $f(t_1, \dots, t_n)$ is determined by an \emph{effect function} $\epsilon : \category{M}^n \to \category{M}$ associated to $f$.
Although the latter type of operations is also important to give natural presentations of computational effects, we leave it for future work.

\paragraph{Enriched Lawvere theories.}
There are many variants of Lawvere theories~\cite{DBLP:journals/entcs/Power06a,HYLAND2007437,EnrichedLawvere,HYLAND2006144,NISHIZAWA2009377,lack_power_2009,BERGER20122029,SystemOfArities,STATON2014197}, and most of them share a common pattern: they are defined as an identity-on-objects functor from a certain category (e.g., $\aleph_0^{\op}$) which represents arities, and the functor must preserve a certain class of products (or cotensors if enriched).
Among the most relevant work to ours are enriched Lawvere theories~\cite{EnrichedLawvere} and discrete Lawvere theories~\cite{HYLAND2006144}.

For a given monoidal category $\category{V}$, a Lawvere $\category{V}$-theory is defined as an identity-on-objects finite cotensor (i.e.\ $\category{V}_{\mathrm{fp}}$-cotensor) preserving $\twist{\category{V}}$-functor $J : \category{V}_{\mathrm{fp}}^{\op} \to \category{L}$ where $\category{V}_{\mathrm{fp}}$ is the full subcategory of $\category{V}$ spanned by finitely presentable objects.
If $\category{V} = \twist{\ordMSet{M}}$, Lawvere $\twist{\ordMSet{M}}$-theories are analogous to our graded Lawvere theories except that we used $\initialGLaw{\category{M}}$ instead of $(\ordMSet{M})_{\mathrm{fp}}$.
Since $n \cdot y(\identity) \in \initialGLaw{\category{M}}$ is finitely presentable, we can say that the notion of graded Lawvere theory is obtained from enriched Lawvere theories by restricting arities to $\initialGLaw{\category{M}} \subseteq (\ordMSet{M})_{\mathrm{fp}}$.
However, the correspondence to finitary graded monads on $\Set$ is an interesting point of our graded Lawvere theories compared to Lawvere $\category{V}$-theories, which correspond to finitary $\category{V}$-monads on $\category{V}$.

Discrete Lawvere theories restrict arities of Lawvere $\category{V}$-theories to $\aleph_0$, that is, a discrete Lawvere $\category{V}$-theory is defined as a ($\Set$-enriched) finite-product preserving functor $J : \aleph_0^{\op} \to \category{L}_0$ where $\category{L}$ is a $\twist{\category{V}}$-category.
Actually, discrete Lawvere $\twist{\ordMSet{M}}$-theories are equivalent to graded Lawvere theories because there is a finite-product preserving functor $\iota : \aleph_0^{\op} \to \initialGLaw{\category{M}}$ such that the composition with $\iota$ gives a bijection between graded Lawvere theories $J : \initialGLaw{\category{M}} \to \category{L}$ and discrete Lawvere $\twist{\ordMSet{M}}$-theories $J_0 \comp \iota : \aleph_0^{\op} \to \category{L}_0$.
However, we considered not only symmetric monoidal categories but also nonsymmetric ones, which cause a nontrivial problem when we define tensor products of algebraic theories.
The problem is that adding commutation equations requires some kind of commutativity of monoidal categories.
We solved this problem by considering product monoidal categories and defining the tensor product of an $\category{M}_1$-graded theory and an $\category{M}_2$-graded theory as an $\category{M}_1 \times \category{M}_2$-graded theory, and the use of two different monoidal categories is new to the best of our knowledge.

\section{Conclusions and Future Work}\label{sec:conclusion}
To extend the correspondence between algebraic theories, Lawvere theories, and (finitary) monads, we introduced notions of graded algebraic theory and graded Lawvere theory and proved their correspondence with finitary graded monads.
We also provided sums and tensor products for graded algebraic theories, which are natural extensions of those for ordinary algebraic theories.
Since we do not assume monoidal categories to be symmetric, our tensor products are a bit different from the ordinary ones in that this combines two theories graded by (or enriched in) different monoidal categories.
We hope that these results will lead us to apply many kinds of techniques developed for monads to graded monads.

As future work, we are interested in ``change-of-effects'', that is, changing the monoidal category $\category{M}$ in $\category{M}$-graded algebraic theory along a (lax) monoidal functor $F : \category{M} \to \category{M}'$.
The problem already appeared in \S\ref{subsec:tensor_products} to define tensor products, but we want to look for more properties of this operation.
We are also interested in integrating a more general framework for notions of algebraic theory~\cite{fujii2019unified} and obtaining a graded version of the framework.
Another direction is exploiting models of graded algebraic theories as modalities in the study of coalgebraic modal logic~\cite{DBLP:conf/calco/MiliusPS15,DBLP:conf/concur/DorschMS19} or weakest precondition semantics~\cite{DBLP:journals/tcs/Hasuo15}.

\paragraph{Acknowledgement.}
We thank Soichiro Fujii, Shin-ya Katsumata, Yuichi Nishiwaki, Yoshihiko Kakutani and the anonymous referees for useful comments.
This work was supported by JST ERATO HASUO Metamathematics for Systems Design Project (No. JPMJER1603).

\ifthenelse{\boolean{longversion}}{
\clearpage
\appendix
\section{Detailed proofs of \S\ref{sec:equivalence}}\label{sec:appendix:proof_equivalence}
We show detailed proofs omitted in \S\ref{sec:equivalence}.

\subsection{Graded Algebraic Theories and Graded Lawvere Theories}
We define an interpretation of $\theory{T}$ in an $\ordMSet{M}$-category with $\initialGLaw{\category{M}}$-cotensors.
\begin{mydefinition}\label{def:M_Set_interpret}
Let $\category{A}$ be an $\ordMSet{M}$-enriched category with $\initialGLaw{\category{M}}$-cotensors, $X$ be an object of $\category{A}$ and $\alpha$ be a family $(\alpha_{n, m} : \Sigma_{n, m} \to \category{A}(\underline{n} \cotensor X, X)m)_{n, m}$ of function.
We define $|\cdot|^{X, \alpha} : T^{\Sigma}_m(n) \to \category{A}(\underline{n} \cotensor X, X) m$ as follows.
\begin{itemize}
\item If $x_i$ is a variable, $|x_i|^{X, \alpha} = \pi_i$.
\item For each operation $f$, $|f(t_1, \dots, t_n)|^{X, \alpha} = \alpha(f) \comp \langle |t_1|^{X, \alpha}, \dots, |t_n|^{X, \alpha} \rangle$.
\item For each $w : m \to m'$, $|c_w(t)|^{X, \alpha} = \category{A}(\underline{n} \cotensor X, X)w (|t|^{X, \alpha})$.
\end{itemize}
\end{mydefinition}
Actually, if $\category{A}$ is an $\ordMSet{M}$-category defined in Definition~\ref{def:enrichment_by_action} and Lemma~\ref{lem:enrichment_by_action_has_contensors}, then Definition~\ref{def:M_Set_interpret} coincides with Definition~\ref{def:term_interpretation}.

%

The above $|\cdot|^{X, \alpha}$ induces an $\ordMSet{M}$-functor that preserves $\initialGLaw{\category{M}}$-cotensors (i.e.\ a model of graded Lawvere theory).
\begin{mydefinition}
We define an $\ordMSet{M}$-functor $\overline{(X, \alpha)} : \theoryfunctor \theory{T} \to \category{A}$ as follows.
\begin{itemize}
	\item $\overline{(X, \alpha)} n = \underline{n} \cotensor X$
	\item A natural transformation $\overline{(X, \alpha)}_{n, n'} : \theoryfunctor \theory{T}(n, n') \to \category{A}(\underline{n} \cotensor X, \underline{n'} \cotensor X)$ is  defined by $([t_1], \dots [t_{n'}]) \mapsto \langle |t_1|^{X, \alpha}, \dots, |t_{n'}|^{X, \alpha} \rangle$.
\end{itemize}
\end{mydefinition}
\begin{mylemma}\label{lem:interpretation_induces_model_of_theory_functor}
The $\ordMSet{M}$-functor $\overline{(X, \alpha)} : \theoryfunctor \theory{T} \to \category{A}$ preserves $\initialGLaw{\category{M}}$-cotensors.
\end{mylemma}
\begin{proof}
%
	By definition, projections are preserved by $\overline{(X, \alpha)}$.
	\qed
\end{proof}

We define the forgetful functor $U : \GradedLawvere_M \to \GradedSpec_M$ by taking all the morphism $f \in L(n, 1) m$ in $L \in \GradedLawvere_M$ as operations and all the equations that hold in $L$ as equational axioms.

\begin{mydefinition}
An $\ordMSet{M}$-functor $U : \GradedLawvere_M \to \GradedSpec_M$ is defined as follows.
\begin{itemize}
\item For each $\category{L} \in \obj \GradedLawvere_M$, $U \category{L} = (\Sigma, E)$ where $\Sigma_{n, m} = \category{L}(n, 1)m$, $E = \{ (s, t) \mid |s|^\category{L} = |t|^\category{L} \}$ and $|\cdot|^\category{L} : T^{\Sigma}_m(n) \to \category{L}(n, 1) m$ is defined by $|\cdot|^{1, 1}$.
\item Given $G : \category{L} \to \category{L'}$, let $UG : U\category{L} \to U\category{L'}$ be a functor defined by
$UG (f) = [G(f)(x_1, \dots, x_n)]$ for each $f \in \category{L}(n, 1) m$.
The equations in $U\category{L}$ is preserved by $UG$, which can be proved by showing $|s|^{(\freefunctor{U \category{L}'} X, UG)} = [(G |s|^\category{L})(x_1, \dots, x_n)]$ for each $s \in T^{U\category{L}}_m(n)$ by induction.
\end{itemize}
\end{mydefinition}
%

Since a graded Lawvere theory $\category{L}$ is a model of itself, we can apply Lemma~\ref{lem:interpretation_induces_model_of_theory_functor} to $\category{L}$.
Actually, we have a one-to-one correspondence as follows.
\begin{mylemma}\label{lem:GLaw_morphism_repr}
For any morphism $G : \theoryfunctor \theory{T} \to \category{L}$ in $\GradedLawvere_{\category{M}}$, there exists a unique family $\alpha = (\alpha_{n, m} : \Sigma_{n, m} \to \category{L}(n, 1) m)_{n, m}$ such that $G = \overline{(1, \alpha)}$.
\end{mylemma}
\begin{proof}
	We define $\alpha$ by $\alpha_{n, m}(f) \coloneqq G_{n, 1, m}([f(x_1, \dots, x_n)])$ for each $f \in \Sigma_{n, m}$.
	Uniqueness of $\alpha$ can be easily proved.
	\qed
\end{proof}

Now, we prove Lemma~\ref{lem:ThT_universal} and Theorem~\ref{thm:GS_GLaw_equiv}.
\begin{proof}[Proof of Lemma~\ref{lem:ThT_universal}]
Let $\tilde{\alpha}$ be a family of functions $\tilde{\alpha}_{n, m} : \Sigma_{n, m} \to \category{L}(n, 1)m$ defined by $\tilde{\alpha}(f) = |t_f|^\category{L}$ where $t_f \in T^{U\category{L}}_m(n)$ is a term that satisfies $\alpha(f) = [t_f]$.
We have $\alpha(f) = [\tilde{\alpha}(f)(x_1, \dots, x_n)]$ for each $f \in \Sigma_{n, m}$ by the definition of equational axioms in $U \category{L}$.
The morphism defined by $\overline{\alpha} \coloneqq \overline{(1, \tilde{\alpha})}$ satisfies $\alpha = U \overline{\alpha} \comp \eta_{\theory{T}}$.

We prove the uniqueness of $\overline{\alpha}$.
Assume $G : \theoryfunctor \theory{T} \to \category{L}$ satisfies $U G \comp \eta = \alpha$.
By Lemma~\ref{lem:GLaw_morphism_repr}, there exists $\beta$ such that $G = \overline{(1, \beta)}$.
For each $f$, we have
\begin{align*}
[\beta(f)(x_1, \dots, x_n)] = U G \comp \eta (f) = \alpha(f) = [\tilde{\alpha}(f)(x_1, \dots, x_n)].
\end{align*}
Therefore, $G = \overline{(1, \beta)} = \overline{(1, \tilde{\alpha})}$.
\qed
\end{proof}

\begin{proof}[Proof of Theorem~\ref{thm:GS_GLaw_equiv}]
We prove that the unit $\eta : \mathrm{Id} \to U \theoryfunctor{}$ and the counit $\epsilon : \theoryfunctor{U} \to \mathrm{Id}$ are isomorphic.
The inverse of the unit $\eta_{\theory{T}}$ is the family of identity functions $1 : \freefunctor{\theory{T}} n m \to \freefunctor{\theory{T}} n m$.
The inverse of the counit $\epsilon_L = \overline{(1, \tilde{1_{UL}})} : \theoryfunctor U L \to L$ is given by $\epsilon^{-1}(f) = ([(\pi_1 \comp f)(x_1, \dots, x_n)], \dots, [(\pi_n \comp f)(x_1, \dots, x_n)])$.
The rest of the proof is routine.
\qed
\end{proof}

We can also prove the equivalence of the categories of models.
\begin{mylemma}
	If $\category{C}$ is a category satisfying $\category{M}$-model condition, then $\modelcat{\theory{T}}(\category{C})$ is equivalent to $\mathrm{Mod}(\theoryfunctor \theory{T}, \KleisliLike{\category{C}}{T})$ where $T$ is the $\twist{\category{M}}$-action on $\category{C}$.
\end{mylemma}
\begin{proof}
	We define a functor $G : \modelcat{\theory{T}}(\category{C}) \to \mathrm{Mod}(\theoryfunctor \theory{T}, \KleisliLike{\category{C}}{T})$ by
	\begin{itemize}
		\item for each model $(A, |\cdot|^A)$ of $\theory{T}$, $G (A, |\cdot|^A) = \overline{(A, |\cdot|^A)}$ and
		\item for each homomorphism $\alpha : (A, |\cdot|^A) \to (B, |\cdot|^B)$ and $n \in \theoryfunctor \theory{T}$, $(G \alpha)_n = \alpha^n : \overline{(A, |\cdot|^A)} n \to \overline{(B, |\cdot|^B)} n$.
	\end{itemize}
	The functor $G$ is a fully faithful and essentially surjective.
\qed
\end{proof}

\subsection{Graded Lawvere theories and Finitary Graded Monads}
In this section, we prove that the category of graded Lawvere theories $\GradedLawvere_{\category{M}}$ and the category of finitary graded monads $\GradedMonad_{\category{M}}$ are equivalent.
Given a graded Lawvere theory, a finitary graded monad is obtained as a coend that represents the set of terms.
On the other hand, given a finitary graded monad, a graded Lawvere theory is obtained from taking the full sub-$\ordMSet{M}$-category on arities $\obj (\initialGLaw{\category{M}})$ of the opposite category of the Kleisli(-like) category in Definition~\ref{def:enrichment_by_action}.
These constructions give rise to an equivalence of categories.

First, we generalize the tupling operation of morphisms in an $\ordMSet{M}$-category with $\initialGLaw{\category{M}}$-cotensors.
\begin{mydefinition}
Let $\category{C}$ be an $\ordMSet{M}$-category with $\initialGLaw{\category{M}}$-cotensors.
Let $f = \langle f_1, \dots, f_{n_1} \rangle \in \category{C}(X, n_1 \cotensor Y) m$ and $g = \langle g_1, \dots, g_{n_2} \rangle \in \category{C}(X, n_2 \cotensor Y) m$.
We define $\langle f, g \rangle \in \category{C}(X, (n_1 + n_2) \cotensor Y) m$ by $\langle f, g \rangle \coloneqq \langle f_1, \dots, f_{n_1}, g_1, \dots, g_{n_2} \rangle$.
\end{mydefinition}

%

An $\category{M}$-graded Lawvere theory yields a finitary graded monad by letting $m \ast X$ be the set of terms of grade $m$ whose variables range over $X$.
\begin{mydefinition}
Let $\category{L}$ be an $\category{M}$-graded Lawvere theory.
We define a finitary $\category{M}$-graded monad $T_{\category{L}} = (\ast, \eta, \mu)$ as follows.
\begin{itemize}
	\item A functor ${\ast} : \category{M} \times \Set \to \Set$  is defined by
		\[ m \ast X \coloneqq \int^{n \in \aleph_0} \category{L}(\underline{n}, \underline{1})m \times X^n \]
		where we regard $\category{L}(\underline{{-}}, \underline{1})$ as the ($\Set$-)functor $\aleph_0 \to \ordMSet{M}$ defined by $n \mapsto \category{L}(\underline{n}, \underline{1}) m$ for objects and the composite
		\begin{center}
			\begin{tikzcd}
				\Set(n, n') \ar[r] & \initialGLaw{\category{M}}(\underline{n'}, \underline{n}) \identity \ar[r] & \category{L}(\underline{n'}, \underline{n}) \identity \ar[r] & \ordMSet{M}(\category{L}(\underline{n}, \underline{1}), \category{L}(\underline{n'}, \underline{1}))
			\end{tikzcd}
		\end{center}
	for morphisms.
	Let $\kappa_{m, X, n} : \category{L}(n, 1)m \times X^n \to m \ast X$ be the universal extranatural transformation.
	\item The unit $\eta_X : X \to \identity \ast X$ is defined by $\eta_X \coloneqq \kappa_{\identity, X, 1}(1_1, {-})$.
	\item Observe that $m_1 \ast (m_2 \ast X)$ is a coend of the following form.
	\begin{align*}
m_1 \ast (m_2 \ast X) &= \int^{n} L(n, 1) m_1 \times \left( \int^{n'} L(n', 1) m_2 \times X^{n'} \right)^n \\
&\cong \int^{n, n_1, \dots, n_n} L(n, 1) m_1 \times \prod_{i=1}^{n} \big( L(n_i, 1) m_2 \times X^{n_i} \big)
	\end{align*}
	The multiplication $\mu_{m_1, m_2, X} : m_1 \ast (m_2 \ast X) \to (m_1 \mult m_2) \ast X$ is a unique morphism induced by the universal property of the coend and the extranatural transformation $\mu'$ below.
	\begin{gather*}
\mu' : L(n, 1) m_1 \times \prod_{i=1}^{n} \big( L(n_i, 1) m_2 \times X^{n_i} \to (m_1 \mult m_2) \ast X \\
\mu'(f, \{ (g_i, x_i) \}_{i}) = \kappa_{m_1 \mult m_2, X, \sum_{i=1}^{n} n_i}(f \comp \langle g_1, \dots, g_n \rangle, (x_1, \dots, x_n))
	\end{gather*}
\end{itemize}
\end{mydefinition}
Note that for each $n' \in \aleph_0$, we have the following isomorphism.
\[ m \ast n' = \int^{n \in \aleph_0} \category{L}(\underline{n}, \underline{1})m \times \Set(n, n') \cong \left(\category{L}(\underline{n}, \underline{1})m\right)^{n'} \cong \category{L}(\underline{n}, \underline{n'})m \]

Conversely, given a graded monad, a graded Lawvere theory is obtained as follows.

\begin{mydefinition}
Let $T = (\ast, \eta, \mu)$ be an $\category{M}$-graded monad on $\Set$.
Let $\category{L}_T$ be an $[\category{M}, \Set]$-category defined as a full sub-$[\category{M}, \Set]$-category of $(\KleisliLike{\Set}{T})^{\op}$ with $\obj(\category{L}_T) = \nat$.
\end{mydefinition}
Since $\category{L}_T$ has $\category{N}_M$-cotensors $\underline{n} \cotensor 1 = n$ whose projections are given by $\pi_i = (() \mapsto \eta(i)) \in \Set(1, \identity * n)$, $\category{L}_T$ is a graded Lawvere theory.

In the rest of this section, to distinguish from the composition in $\Set$, the composition in $\category{L}_T$ is denoted by $\odot$, that is, for each $f \in \category{L}_T(n', n'') m_1$ and $g \in \category{L}_T(n, n') m_2$, $f \odot g = \mu \comp (m_1 \ast g) \comp f$.

\begin{mylemma}\label{lem:GLawGMndGLaw}
	$\category{L}_{T_\category{L}} \cong \category{L}$
\end{mylemma}
\begin{proof}
	The following isomorphism between hom-objects induces an isomorphism $\category{L}_{T_\category{L}} \cong \category{L}$.
	\[ \category{L}_{T_{\category{L}}}(n, n') m = \Set(n', m \ast n) \cong (\category{L}(n, 1) m)^{n'} \cong \category{L}(n, n') m \]
	\qed
\end{proof}


Given an morphism $\alpha : T \to T'$ in $\GradedMonad_{\category{M}}$, we define $\category{L}_{\alpha} : \category{L}_T \to \category{L}_{T'}$ by $(\category{L}_{\alpha})_{n, n', m} = \Set(n', \alpha_{n, m}) : \category{L}_T(n, n') m \to \category{L}_{T'}(n, n') m$.
It is easy to prove that $\category{L}_{\alpha}$ is a morphism in $\GradedLawvere_{\category{M}}$ and $\category{L}_{({-})} : \GradedMonad_{\category{M}} \to \GradedLawvere_{\category{M}}$ is a functor.

We prove Theorem~\ref{thm:GLaw_GMnd_equiv}.
\begin{proof}[Proof of Theorem~\ref{thm:GLaw_GMnd_equiv}]
We prove that $\category{L}_{({-})} : \GradedMonad_{\category{M}} \to \GradedLawvere_{\category{M}}$ is an essentially surjective fully faithful functor.

Let $\alpha, \beta : T \to T'$ be morphisms between finitary graded monads and assume $\category{L}_{\alpha} = \category{L}_{\beta}$.
To prove $\alpha = \beta$, it suffices to prove $\alpha_{m, n} = \beta_{m, n}$ for each $m \in \category{M}$ and $n \in \aleph_0$ by the finitaryness of $T$ and $T'$, which follows from $\Set(1, \alpha_{m, n}) = (\category{L}_{\alpha})_{n, 1, m} = (\category{L}_{\beta})_{n, 1, m} = \Set(1, \beta_{m, n})$.
Therefore, $\category{L}_{({-})}$ is faithful.

Let $T = (\ast, \eta, \mu)$ and $T' = (\ast', \eta', \mu')$ be finitary graded monads.
We prove that for each $G : \category{L}_{T} \to \category{L}_{T'}$, there exists $\alpha : T \to T'$ such that $G = \category{L}_{\alpha}$.
For each $m \in \category{M}$ and $n \in \aleph_0$, we define $\alpha_{m, n} : m \ast n \to m \ast' n$ as the unique function satisfying $G_{n, 1, m} = \Set(1, \alpha_{m, n}) : \Set(1, m \ast n) \to \Set(1, m \ast' n)$.
This $\alpha_{m, n}$ satisfies the conditions for $\eta$ and $\mu$ in Lemma~\ref{lem:finitary_graded_monad_morphism}, and we have a morphism $\alpha : T \to T'$ induced by $\alpha_{m, n}$.
This $\alpha$ satisfies $\category{L}_{\alpha} = G$.
Therefore, $\category{L}_{({-})}$ is full.

$\category{L}_{({-})}$ is essentially surjective by Lemma~\ref{lem:GLawGMndGLaw}.
\qed
\end{proof}

}{}

\bibliographystyle{splncs04}
\bibliography{ref}


\vfill

{\small\medskip\noindent{\bf Open Access} This chapter is licensed under the terms of the Creative Commons\break Attribution 4.0 International License (\url{http://creativecommons.org/licenses/by/4.0/}), which permits use, sharing, adaptation, distribution and reproduction in any medium or format, as long as you give appropriate credit to the original author(s) and the source, provide a link to the Creative Commons license and indicate if changes were made.}

{\small \spaceskip .28em plus .1em minus .1em The images or other third party material in this chapter are included in the chapter's Creative Commons license, unless indicated otherwise in a credit line to the material.~If material is not included in the chapter's Creative Commons license and your intended\break use is not permitted by statutory regulation or exceeds the permitted use, you will need to obtain permission directly from the copyright holder.}

\medskip\noindent\includegraphics{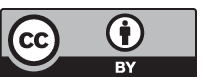}

\end{document}